\documentclass[a4paper,10pt]{article}
\usepackage{amssymb}
\usepackage{amsmath}
\usepackage{graphicx}
\usepackage{dcolumn}
\usepackage{bm}
\usepackage{lineno}
\usepackage{subfigure}
\usepackage{enumerate}
\usepackage{float}
\usepackage[hmargin={1.5cm,1.5cm}, vmargin={2cm,2cm}]{geometry}
\usepackage[dvipdfm,colorlinks,citecolor=blue,linkcolor=blue]{hyperref}
\graphicspath{{./}{./figs/}}

\numberwithin{equation}{section}

\begin{document}

\title{Signal Propagation in Feedforward Neuronal Networks with Unreliable Synapses}

\author{Daqing Guo$^1$\thanks{Email: dqguo07@gmail.com}, \ \ \ Chunguang Li$^2$\thanks{Author for correspondence, Email: cgli@zju.edu.cn}}
\date{\small $^1$ School of Electronic Engineering, University of Electronic Science and Technology of China, Chengdu 610054, People's Republic of China \\ $^2$ Department of Information Science and Electronic Engineering, Zhejiang University, Hangzhou 310027, People's Republic of China \\ \today}

\maketitle

\begin{abstract}
In this paper, we systematically investigate both the synfire
propagation and firing rate propagation in feedforward neuronal
network coupled in an all-to-all fashion. In contrast to most
earlier work, where only reliable synaptic connections are
considered, we mainly examine the effects of unreliable
synapses on both types of neural activity propagation in this
work. We first study networks composed of purely excitatory
neurons. Our results show that both the successful transmission
probability and excitatory synaptic strength largely influence
the propagation of these two types of neural activities, and
better tuning of these synaptic parameters makes the considered
network support stable signal propagation. It is also found
that noise has significant but different impacts on these two
types of propagation. The additive Gaussian white noise has the
tendency to reduce the precision of the synfire activity,
whereas noise with appropriate intensity can enhance the
performance of firing rate propagation. Further simulations
indicate that the propagation dynamics of the considered
neuronal network is not simply determined by the average amount
of received neurotransmitter for each neuron in a time instant,
but also largely influenced by the stochastic effect of
neurotransmitter release. Second, we compare our results with
those obtained in corresponding feedforward neuronal networks
connected with reliable synapses but in a random coupling
fashion. We confirm that some differences can be observed in
these two different feedforward neuronal network models.
Finally, we study the signal propagation in feedforward
neuronal networks consisting of both excitatory and inhibitory
neurons, and demonstrate that inhibition also plays an
important role in signal propagation in the considered
networks.
\begin{flushleft}
\end{flushleft}
\end{abstract}


\newpage

\section{Introduction}
\label{sec:1}

A major challenge in neuroscience is to understand how the
neural activities are propagated through different brain
regions, since many cognitive tasks are believed to involve
this process (Vogels and Abbott, 2005). The feedforward
neuronal network is the most used model in investigating this
issue, because it is simple enough yet can explain propagation
activities observed in experiments. In recent years, two
different modes of neural activity propagation have been
intensively studied. It has been found that both the
synchronous spike packet (\textit{synfire}), and the
\textit{firing rate}, can be transmitted across deeply layered
networks (Abeles 1991; Aertsen et al. 1996; Diesmann et al.
1999; Diesmann et al. 2001; C$\hat{\text{a}}$teau and Fukai
2001; Gewaltig et al. 2001; Tetzlaff et al. 2002; Tetzlaff et
al. 2003; van Rossum et al. 2002; Vogels and Abbott 2005; Wang
et al. 2006; Aviel et al. 2003; Kumar et al. 2008; Kumar et al.
2010; Shinozaki et al. 2007; Shinozaki et al. 2010). Although
these two propagation modes are quite different, the previous
results demonstrated that a single network with different
system parameters can support stable and robust signal
propagation in both of the two modes, for example, they can be
bridged by the background noise and synaptic strength (van
Rossum et al. 2002; Masuda and Aihara 2002; Masuda and Aihara
2003).

Neurons and synapses are fundamental components of the brain.
By sensing outside signals, neurons continually fire discrete
electrical signals known as action potentials or so-called
spikes, and then transmit them to postsynaptic neurons through
synapses (Dayan and Abbott 2001). The spike generating
mechanism of cortical neurons is generally highly reliable.
However, many studies have shown that the communication between
neurons is, by contrast, more or less unreliable (Abeles 1991;
Raastad et al. 1992; Smetters and Zador 1996). Theoretically,
the synaptic unreliability can be explained by the phenomenon
of probabilistic transmitter release (Branco and Staras 2009;
Katz 1966; Katz 1969; Trommersh\"{a}user et al. 1999), i.e.,
synapses release neurotransmitter in a stochastic fashion,
which has been confirmed by well-designed biological
experiments (Allen and Stevens 1994). In most cases, the
transmission failure rate at a given synapse tends to exceed
the fraction of successful transmission (Rosenmund et al. 1993;
Stevens and Wang 1995). In some special cases, the synaptic
transmission failure rate can be as high as 0.9 or even higher
(Allen and Stevens 1994). Further computational studies have
revealed that the unreliability of synaptic transmission might
be a part of information processing of the brain and possibly
has functional roles in neural computation. For instance, it
has been reported that the unreliable synapses provide a useful
mechanism for reliable analog computation in space-rate coding
(Maass and Natschl$\ddot{\text{a}}$ger 2000); and it has been
found that suitable synaptic successful transmission
probability can improve the information transmission efficiency
of synapses (Goldman 2004) and can filter the redundancy
information by removing autocorrelations in spike trains
(Goldman et al. 2002). Furthermore, it has also been
demonstrated that unreliable synapses largely influence both
the emergence and dynamical behaviors of clusters in an
all-to-all pulse-coupled neuronal network, and can make the
whole network relax to clusters of identical size (Friedrich
and Kinzel 2009).

Although the signal propagation in multilayered feedforward
neuronal networks has been extensively studied, to the best of
our knowledge the effects of unreliable synapses on the
propagation of neural activity have not been widely discussed
and the relevant questions still remain unclear (but see the
footnote\footnote{An anonymous reviewer kindly reminded us that
there might be a relevant abstract (Trommersh\"{a}user and
Diesmann 2001) discussing the effect of synaptic variability on
the synchronization dynamics in feedforward cortical neural
networks, but the abstract itself does not contain the results
presumably presented on the poster and also the follow-up
publications do not exist.}). In this paper, we address these
questions and provide insights by computational modeling. For
this purpose, we examine both the synfire propagation and
firing rate propagation in feedforward neuronal networks. We
mainly investigate the signal propagation in feedforward
neuronal networks composed of purely excitatory neurons
connected with unreliable synapses in an all-to-all coupling
fashion (abbr. URE feedforward neuronal network) in this work.
We also compare our results with the corresponding feedforward
neuronal networks (we will clarify the meaning of
``corresponding'' later) composed of purely excitatory neurons
connected with reliable synapses in a random coupling fashion
(abbr. RRE feedforward neuronal network). Moreover, we study
feedforward neuronal networks consisting of both excitatory and
inhibitory neurons connected with unreliable synapses in an
all-to-all coupling fashion (abbr. UREI feedforward neuronal
network).

The rest of this paper is organized as follows. The network
architecture, neuron model, and synapse model used in this paper
are described in Sec.~\ref{sec:2}. Besides these, the measures to
evaluate the performance of synfire propagation and firing rate
propagation, as well as the numerical simulation method are also
introduced in this section. The main results of the present work
are presented in Sec.~\ref{sec:3}. Finally, a detailed conclusion
and discussion of our work are given in Sec.~\ref{sec:4}.

\section{Model and method}
\label{sec:2}
\subsection{Network architecture}
\label{sec:2a}

\begin{figure}[!t]
\centering \includegraphics[width=15cm]{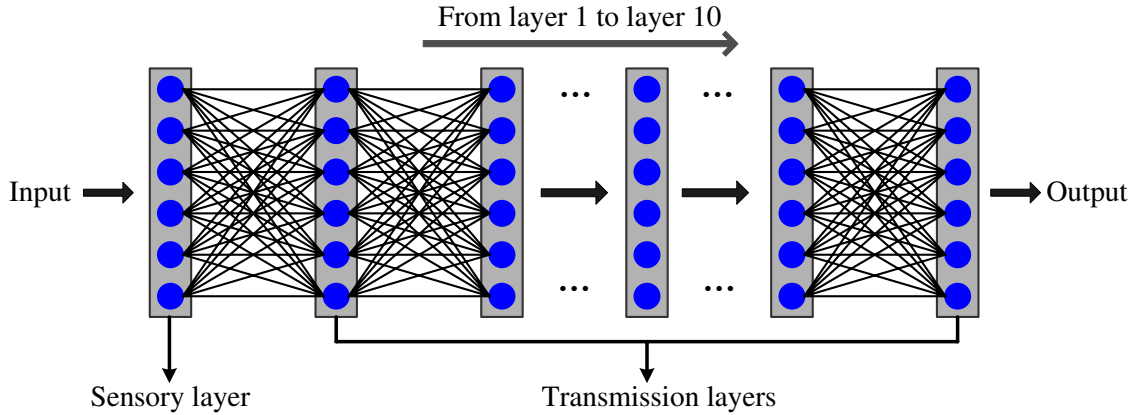}
\caption{\label{fig:1} Network architecture of the URE feedforward
neuronal network. The network totally contains 10 layers. The
first layer is the sensory layer and the others are the
transmission layers. Each layer consists of 100 excitatory
neurons. For clarity, only 6 neurons are shown in each layer.}
\end{figure}

In this subsection, we introduce the network topology used in
this paper. Here we only describe how to construct the URE
feedforward neuronal network. The methods about how to build
the corresponding RRE feedforward neuronal network and the UREI
feedforward neuronal network will be briefly given in
Secs.~\ref{sec:3d} and \ref{sec:3e}, respectively. The
architecture of the URE feedforward neuronal network is
schematically shown in Figure~\ref{fig:1}. The network totally
contains $L=10$ layers, and each layer is composed of
$N_{s}=100$ excitatory neurons. Since neurons in the first
layer are responsible for receiving and encoding the external
input signal, we therefore call this layer sensory layer and
neurons in this layer are called sensory neurons. In contrast,
the function of neurons in the other layers is to propagate
neural activities. Based on this reason, we call these layers
transmission layers and the corresponding neurons cortical
neurons. Because the considered neuronal network is purely
feedforward, there is no feedback connection from neurons in
downstream layers to neurons in upstream layers, and there is
also no connection among neurons within the same layer. For
simplicity, we call the $i$-th neuron in the $j$-th layer
neuron $(i,j)$ in the following.

\subsection{Neuron model}
\label{sec:2b}

We now introduce the neuron model used in the present work.
Each cortical neuron is modeled by using the integrate-and-fire
(IF) model (Nordlie et al. 2009), which is a minimal spiking
neuron model to mimic the action potential firing dynamics of
biological neurons. The subthreshold dynamics of a single IF
neuron obeys the following differential equation:
\begin{equation}
\begin{split}
\tau_m\frac{dV_{ij}}{dt}=V_{\text{rest}}-V_{ij}+RI_{ij},
 \end{split}
 \label{eq:1}
\end{equation}
with the total input current
\begin{equation}
\begin{split}
I_{ij}=I_{ij}^{\text{syn}}+I_{ij}^{\text{noise}}.
\end{split}
\label{eq:2}
\end{equation}
Here $i=1,2,\ldots,N_{s}$ and $j=2,3,\ldots,L$, $V_{ij}$
represents the membrane potential of neuron $(i,j)$, $\tau_m=20$
ms is the membrane time constant, $V_{\text{rest}}=-60$ mV is the
resting membrane potential, $R=20$ M$\Omega$ denotes the membrane
resistance, and $I_{ij}^{\text{syn}}$ is the total synaptic
current. The noise current
$I_{ij}^{\text{noise}}=\sqrt{2D_t}\xi_{ij}(t)$ represents the
external or intrinsic fluctuations of the neuron, where
$\xi_{ij}(t)$ is a Gaussian white noise with zero mean and unit
variance, and $D_t$ is referred to as the noise intensity of the
cortical neurons. In this work, a deterministic threshold-reset
mechanism is implemented for spike generation. Whenever the
membrane potential of a neuron reaches a fixed threshold at
$V_{\text{th}}=-50$ mV, the neuron fires a spike, and then the
membrane potential is reset according to the resting potential,
where it remains clamped for a 5-ms refractory period.

On the other hand, we use different models to simulate the
sensory neurons depending on different tasks. To study the
synfire propagation, we assume that each sensory neuron is a
simple spike generator, and control their firing behaviors by
ourselves. While studying the firing rate propagation, the
sensory neuron is modeled by using the IF neuron model with the
same expression (see Eq.~(\ref{eq:1})) and the same parameter
settings as those for cortical neurons. For each sensory
neuron, the total input current is given by
\begin{equation}
\begin{split}
I_{i1}=I(t)+I_{i1}^{\text{noise}},
\end{split}
\label{eq:3}
\end{equation}
where $i=1,2,...,N_s$ index neurons. The noise current
$I_{i1}^{\text{noise}}$ has the same form as that for cortical
neurons but with the noise intensity $D_s$. $I(t)$ is a
time-varying external input current which is injected to all
sensory neurons. For each run of the simulation, the external
input current is constructed by the following process. Let
$\eta(t)$ denote an Ornstein-Uhlenbeck process, which is
described by
\begin{equation}
\begin{split}
\tau_c\frac{d\eta(t)}{dt}= -\eta(t) +
\sqrt{2A}\xi(t),
\end{split}
\label{eq:4}
\end{equation}
where $\xi(t)$ is a Gaussian white noise with zero mean and unit
variance, $\tau_c$ is a correlation time constant, and $A$ is a
diffusion coefficient. The external input current $I(t)$ is
defined as
\begin{equation}
\begin{split}
I(t)=
\begin{cases}
\eta(t)& \text{if $\eta(t)\geq0$},\\
0& \text{if $\eta(t)<0$}.
\end{cases}
\end{split}
\label{eq:5}
\end{equation}
Parameter $A$ can be used to denote the intensity of the
external input signal $I(t)$. In this work, we choose $A=200$
$\text{nA}^2$ and $\tau_c=80$ ms. By its definition, the
external input current $I(t)$ corresponds to a
Gaussian-distributed white noise low-pass filtered at 80 ms and
half-wave rectified. It should be noted that this type of
external input current is widely used in the literature, in
particular in the research papers which study the firing rate
propagation (van Rossum et al. 2002; Vogels and Abbott 2005;
Wang and Zhou 2009).

\subsection{Synapse model}
\label{sec:2c}

The synaptic interactions between neurons are implemented by
using the modified conductance-based model. Our modeling
methodology is inspired by the phenomenon of probabilistic
transmitter release of the real biological synapses. Here we
only introduce the model of unreliable excitatory synapses,
because the propagation of neural activity is mainly examined
in URE feedforward neuronal networks in this work. The methods
about how to model reliable excitatory synapses and unreliable
inhibitory synapses will be briefly introduced in
Secs.~\ref{sec:3d} and \ref{sec:3e}, respectively.

The total synaptic current onto neuron $(i,j)$ is the linear
sum of the currents from all incoming synapses,
\begin{equation}
\begin{split}
I_{ij}^{\text{syn}}=\sum_{k=1}^{N_s}G(i,j;k,j-1)\cdot(E_{{\text{syn}}}-V_{ij}).
\end{split}
\label{eq:6}
\end{equation}
In this equation, the outer sum runs over all synapses onto
this particular neuron, $G(i,j;k,j-1)$ is the conductance from
neuron $(k,j-1)$ to neuron $(i,j)$, and $E_{\text{syn}}=0$ mV
is the reversal potential of the excitatory synapse. Whenever
the neuron $(k,j-1)$ emits a spike, an increment is assigned to
the corresponding synaptic conductances according to the
synaptic reliability parameter, which process is given by
\begin{equation}
\begin{split}
G(i,j;k,j-1)\leftarrow G(i,j;k,j-1)+J(i,j;k,j-1)\cdot h(i,j;k,j-1),
\end{split}
\label{eq:7}
\end{equation}
where $h(i,j;k,j-1)$ denotes the synaptic reliability parameter
of the synapse from neuron $(k,j-1)$ to neuron $(i,j)$, and
$J(i,j;k,j-1)$ stands for the relative peak conductance of this
particular excitatory synapse which is used to determine its
strength. For simplicity, we assume that $J(i,j;k,j-1)=g$, that
is, the synaptic strength is identical for all excitatory
connections. Parameter $p$ is defined as the successful
transmission probability of spikes. When a presynaptic neuron
$(k,j-1)$ fires a spike, we let the corresponding synaptic
reliability variables $h(i,j;k,j-1)=1$ with probability $p$ and
$h(i,j;k,j-1)=0$ with probability $1-p$. That is to say,
whether the neurotransmitter is successfully released or not is
in essence controlled by a Bernoulli on-off process in the
present work. In other time, the synaptic conductance decays by
an exponential law:
\begin{equation}
\begin{split}
\frac{d}{dt}G(i,j;k,j-1)=\frac{1}{\tau_{s}}G(i,j;k,j-1),
\end{split}
\label{eq:8}
\end{equation}
with a fixed synaptic time constant $\tau_{s}$. In the case of
synfire propagation, we choose $\tau_s=2$ ms, and in the case
of firing rate propagation, we choose $\tau_s=5$ ms.

\subsection{Measures of the synfire and firing rate propagation}
\label{sec:2d}

We now introduce several useful measures used to quantitatively
evaluate the performance of the two different propagation
modes: the synfire mode and firing rate mode. The propagation
of synfire activity is measured by the survival rate and the
standard deviation of the spiking times of the synfire packet
(Gewaltig et al. 2001). Let us first introduce how to calculate
the survival rate for the synfire propagation. In our
simulations, we find that the synfire propagation can be
divided into three types: the failed synfire propagation, the
stable synfire propagation, as well as the synfire instability
propagation (for detail, see Sec.~\ref{sec:3a}). For neurons in
each layer, a threshold method is developed to detect the local
highest ``energy'' region. To this end, we use a 5 ms moving
time window with 0.1 ms sliding step to count the number of
spikes within each window. Here a high energy region means that
the number of spikes within the window is larger than a
threshold $\theta=50$. Since we use a moving time window with
small sliding step, there might be a continuous series of
windows contain more than 50 spikes around a group of
synchronous spikes. In this work, we only select the first
window which covers the largest number of spikes around a group
of synchronous spikes as the local highest energy region. We
use the number of local highest energy region to determine
which type of synfire propagation occurs. If there is no local
highest energy region detected in the final layer of the
network, we consider it as the failed synfire propagation. When
two or more separated local highest energy regions are detected
in one layer, we consider it as the synfire instability
propagation. Otherwise, it means the occurrence of the stable
synfire propagation. For each experimental setting, we carry
out the simulation many times. The survival rate of the synfire
propagation is defined as the ratio of the number of occurrence
of the stable synfire propagation to the total number of
simulations. In additional simulations, it turns out that the
threshold value $\theta$ can vary in a wide range without
altering the results. Under certain conditions, noise can help
the feedforward neuronal network produce the spontaneous spike
packets, which promotes the occurrence of synfire instability
propagation and therefore decreases the survival rate. For
stable synfire propagation, there exists only one highest
energy region for neurons in each layer. Spikes within this
region are considered as the candidate synfire packet, which
might also contain a few spontaneous spikes caused by noise and
other factors. In this work, an adaptive algorithm is
introduced to eliminate spontaneous spikes from the candidate
synfire packet. Suppose now that there is a candidate synfire
packet in the $i$-th layer with the number of spikes it
contains $\alpha_i$ and the corresponding spiking times
$\{t_1,t_2,\ldots,t_{\alpha_i}\}$. The average spiking time of
the candidate synfire packet is therefore given by
\begin{equation}
\begin{split}
\bar{t}_i=\frac{1}{\alpha_i}\sum_{k=1}^{\alpha_i}t_k.
\end{split}
\label{eq:9}
\end{equation}
Thus the standard deviation of the spiking times in the $i$-th
layer can be calculated as follows:
\begin{equation}
\begin{split}
\sigma_i=\sqrt{\frac{1}{\alpha_i}\sum_{k=1}^{\alpha_i}[t_k-\bar{t}_i]^2}.
\end{split}
\label{eq:10}
\end{equation}
We remove the $j$-th spike from the candidate synfire packet if
it satisfies: $|t_j-\bar{t}_i|>\mu\sigma_i$, where $\mu$ is a
parameter of our algorithm. We recompute the average spiking
time as well as the standard deviation of the spiking times for
the new candidate synfire packet, and repeat the above
eliminating process, until no spike is removed from the new
candidate synfire packet anymore. We define the remaining
spikes as the synfire packet, which is characterized by the
final values of $\alpha_i$ and $\sigma_i$. Parameter $\mu$
determines the performance of the proposed algorithm. If $\mu$
is too large, the synfire packet will lose several useful
spikes at its borders, and if $\mu$ is too small, the synfire
packet will contain some noise data. In our simulations, we
found that $\mu=4$ can result in a good compromise between
these two extremes. It should be emphasized that our algorithm
is inspired by the method given in (Gewaltig et al. 2001).
Next, we introduce how to measure the performance of the firing
rate propagation. The performance of firing rate propagation is
evaluated by combining it with a population code. Specifically,
we compute how similar the population firing rates in different
layers to the external input current $I(t)$ (van Rossum et al.
2002; Vogels and Abbott 2005). To do this, a 5 ms moving time
window with 1 ms sliding step is also used to estimate the
population firing rates ${r_i(t)}$ for different layers as well
as the smooth version of the external input current $I_s(t)$.
The correlation coefficient between the population firing rate
of the $i$-th layer and external input current is calculated by
\begin{equation}
\begin{split}
C_i(\tau) =
\frac{\left\langle\left[I_s(k+\tau)-\overline{I}_s\right]
\left[r_i(k)-\overline{r}_i\right]\right\rangle_t}{\sqrt{\left\langle
\left[I_s(k+\tau)-\overline{I}_s\right]^2\right\rangle_t\left
\langle\left[r_i(k)-\overline{r}_i\right]^2\right\rangle_t}},
 \end{split}
  \label{eq:11}
\end{equation}
where $\langle\cdot\rangle_t$ denotes the average over time.
Here we use the maximum cross-correlation coefficient $Q_i=
\max\{C_i(\tau)\}$ to quantify the performance of the firing
rate propagation in the $i$-th layer. Note that $Q_i$ is a
normalization measure and a larger value corresponds to a
better performance.

\subsection{Numerical simulation method}
\label{sec:2e}

In all numerical simulations, we use the standard Euler-Maruyama
integration scheme to numerically calculate the aforementioned
stochastic differential Eqs.~(\ref{eq:1})-(\ref{eq:8}) (Kloeden et
al. 1994). The temporal resolution of integration is fixed at 0.02
ms for calculating the measures of the synfire mode and at 0.05 ms
for calculating the measures of the firing rate mode, as the
measurement of the synfire needs higher precise. In additional
simulations, we have found that further reducing the integration
time step does not change our numerical results in a significant
way. For the synfire mode, all simulations are executed at least
100 ms to ensure that the synfire packet can be successfully
propagated to the final layer of the considered network. While
studying the firing rate mode, we perform all simulations up to
5000 ms to collect enough spikes for statistical analysis. It
should be noted that, to obtain convincing results, we carry out
several times of simulations (at least 200 times for the synfire
mode and 50 times for the firing rate mode) for each experimental
setting to compute the corresponding measures.

\section{Simulation results}
\label{sec:3}

In this section, we report the main results obtained in the
simulation. We first systematically investigate the signal
propagation in the URE feedforward neuronal networks. Then, we
compare these results with those for the corresponding RRE
feedforward neuronal networks. Finally, we further study the
signal propagation in the UREI feedforward neuronal networks.

\subsection{Synfire propagation in URE feedforward neuronal networks}
\label{sec:3a}

\begin{figure}[!t]
\centering
\subfigure{\includegraphics[width=7.8cm]{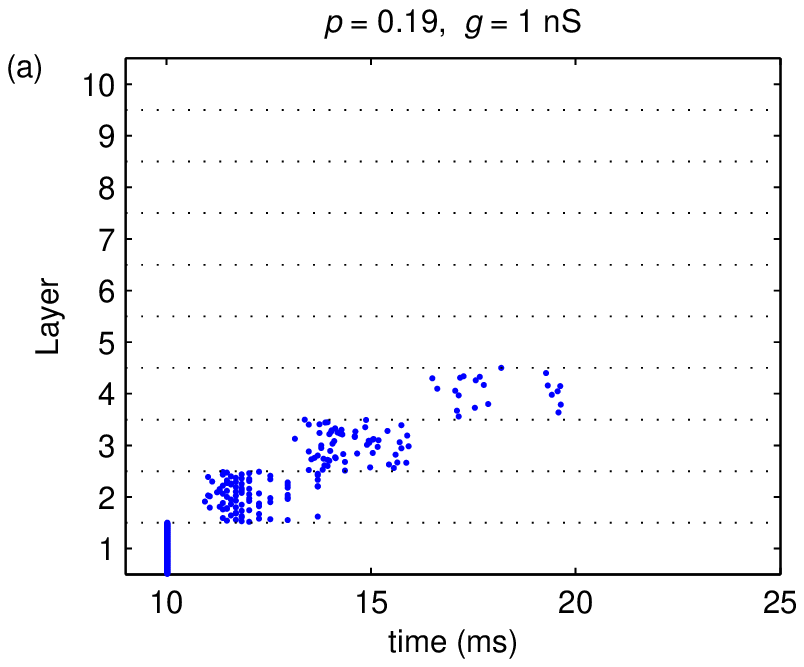}
\label{fig:2a}}
\subfigure{\includegraphics[width=7.8cm]{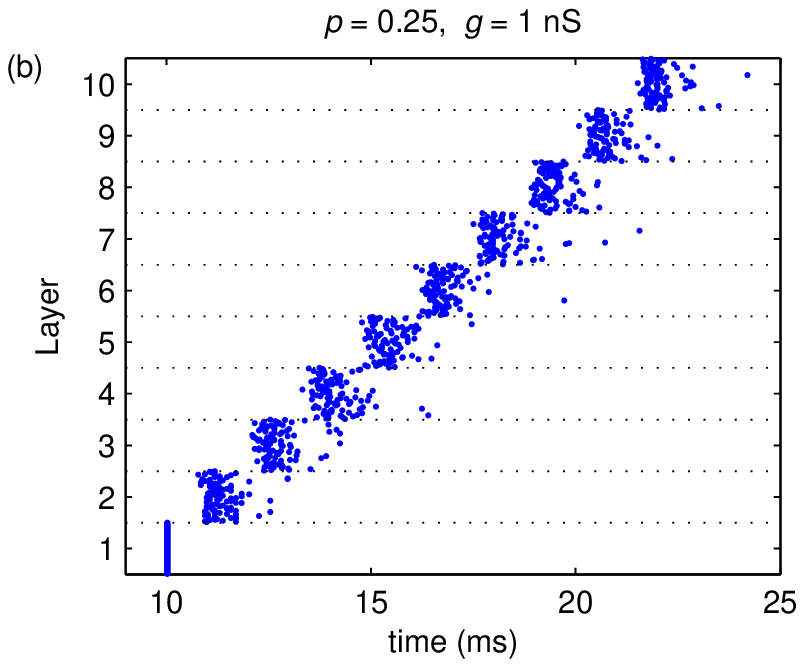} \label{fig:2b}} \\
\subfigure{\includegraphics[width=7.8cm]{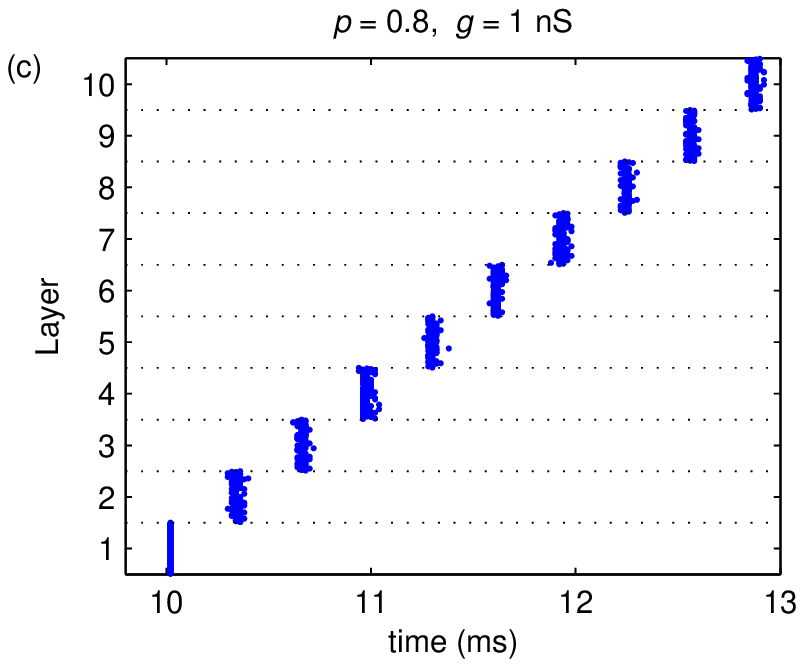} \label{fig:2c}}
\subfigure{\includegraphics[width=7.8cm]{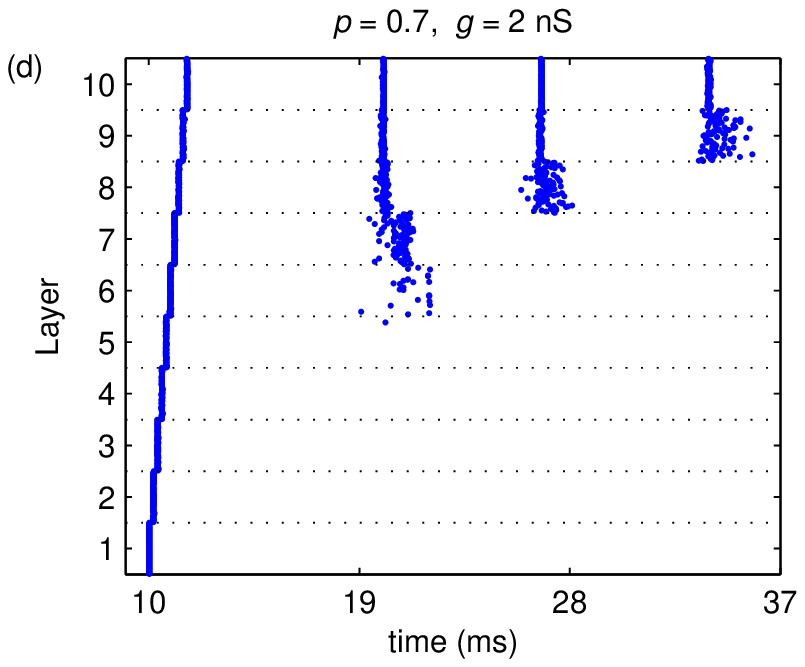} \label{fig:2d}}
\caption{\label{fig:2} (Color online) Several typical spike raster
diagrams for different values of successful transmission
probability $p$ and excitatory synaptic strength $g$. Shown are
samples of (a) failed synfire propagation, (b) and (c) stable
synfire propagation, and (d) synfire instability. System parameters
are $p=0.19$ and $g=1$ nS (a), $p=0.25$ and $g=1$ nS (b), $p=0.8$
and $g=1$ nS (c), and $p=0.7$ and $g=2$ nS (d), respectively. As
we see, the synfire packet can reach the final layer of the
network successfully only for appropriate values of $p$ and $g$.
It should be noted that the time scales in these four subfigures are
different.}
\end{figure}

Here we study the role of unreliable synapses on the
propagation of synfire packet in the URE feedforward neuronal
networks. In the absence of noise, we artificially let each
sensory neuron fire and only fire an action potential at the
same time ($\alpha_1=100$ and $\sigma_1=0$ ms). Without loss of
generality, we let all sensory neurons fire spikes at $t=10$
ms. Figure~\ref{fig:2} shows four typical spike raster diagrams
of propagating synfire activity. Note that the time scales in
Figs.~\ref{fig:2a}-\ref{fig:2d} are different. The URE
feedforward neuronal network with both small successful
transmission probability and small excitatory synaptic strength
badly supports the synfire propagation. In this case, due to
high synaptic unreliability and weak excitatory synaptic
interaction between neurons, the propagation of synfire packet
cannot reach the final layer of the whole network (see
Fig.~\ref{fig:2a}). For suitable values of $p$ and $g$, we find
that the synfire packet can be stably transmitted in the URE
feedforward neuronal network. Moreover, it is obvious that the
width of the synfire packet at any layer for $p=0.8$ is much
narrower than that of the corresponding synfire packet for
$p=0.25$ (see Figs.~\ref{fig:2b} and \ref{fig:2c}). At the same
time, the transmission speed is also enhanced with the
increasing of $p$. These results indicate that the neuronal
response of the considered network is much more precise and
faster for suitable large successful transmission probability.
However, our simulation results also reveal that a strong
excitatory synaptic strength with large value of $p$ might
destroy the propagation of synfire activity. As we see from
Fig.~\ref{fig:2d}, the initial tight synfire packet splits into
several different synfire packets during the transmission
process. Such phenomenon is called the ``synfire instability''
(Tetzlaff et al. 2002; Tetzlaff et al. 2003), which mainly
results from the burst firings of several neurons caused by the
strong excitatory synaptic interaction as well as the
stochastic fluctuation of the synaptic connections.

\begin{figure}[!t]
\centering
\subfigure{\includegraphics[width=7.8cm]{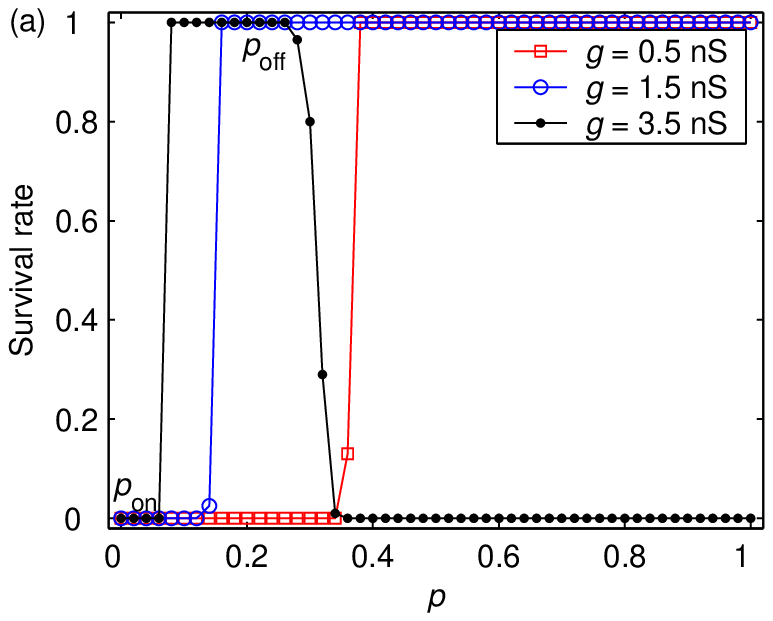}
\label{fig:3a}}
\subfigure{\includegraphics[width=8cm]{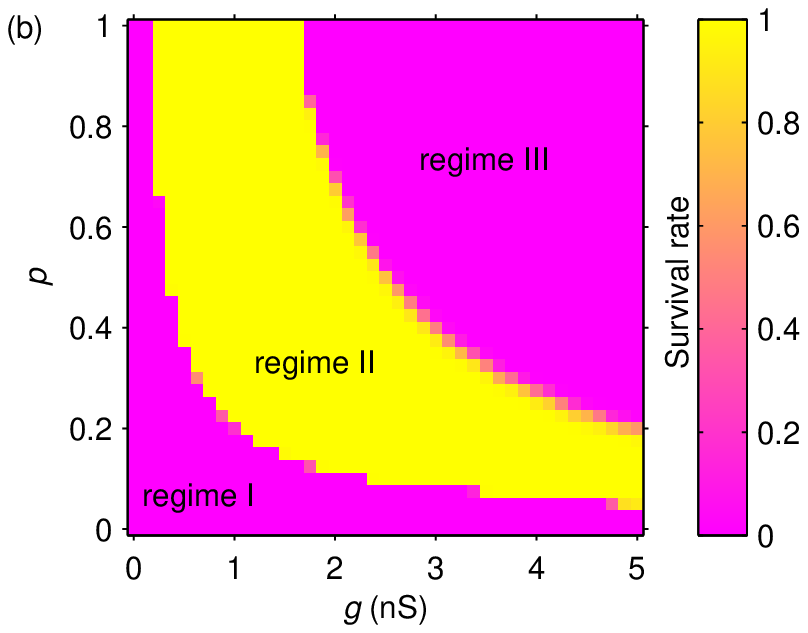} \label{fig:3b}} \\
\subfigure{\includegraphics[width=7.8cm]{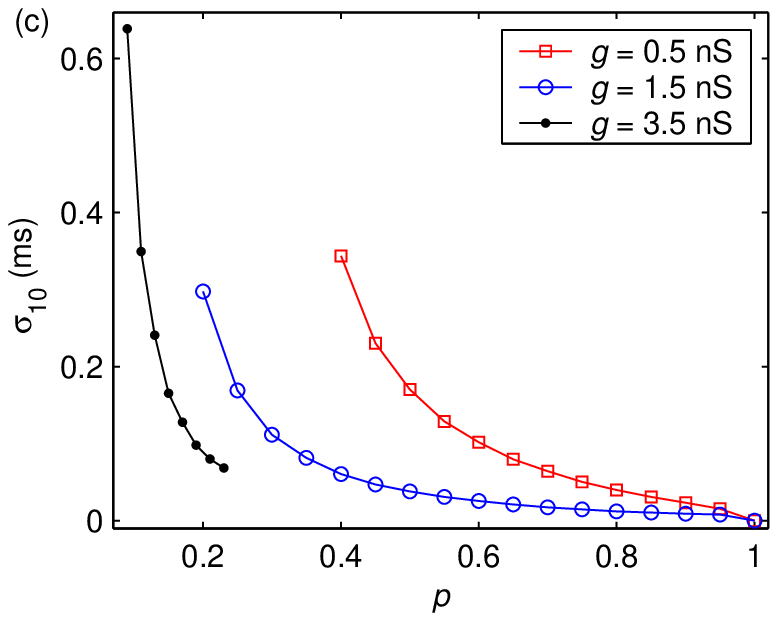} \label{fig:3c}}
\subfigure{\includegraphics[width=7.8cm]{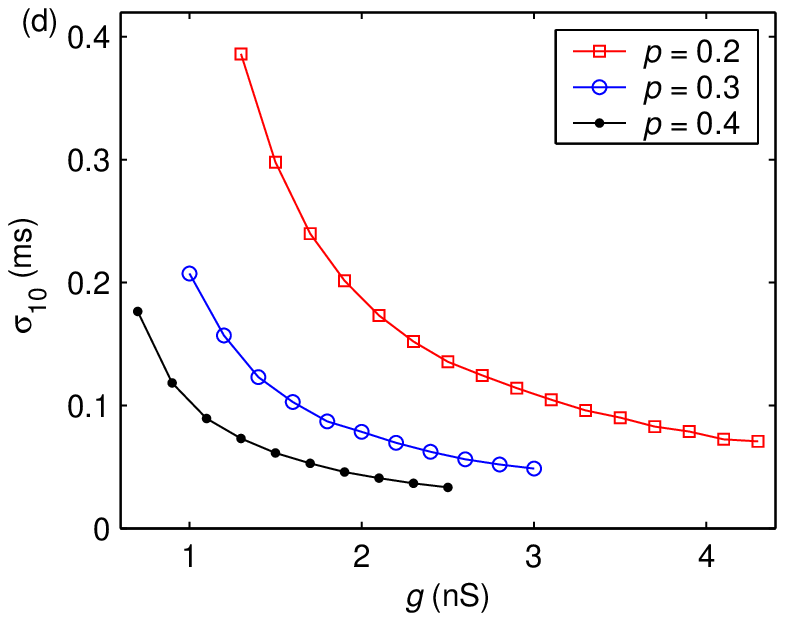} \label{fig:3d}}
\caption{\label{fig:3} (Color online) Effects of the successful
transmission probability and excitatory synaptic strength on the
synfire propagation in URE feedforward neuronal networks. (a) The
survival rate of the synfire propagation versus $p$ for different
values of $g$. (b) Schematic of three different synfire
propagation regimes in the $(g,p)$ panel ($41\times41=1681$
points). Regime I: the failed synfire propagation region; regime
II: the stable synfire propagation region; and regime III: the
synfire instability propagation region. (c) The value of
$\sigma_{10}$ as a function of $p$ for different values of $g$.
(d) The value of $\sigma_{10}$ as a function of $g$ for different
values of $p$. In all cases, the noise intensity $D_t=0$. Each
data point shown here is computed based on 200 independent
simulations with different random seeds.}
\end{figure}

In Fig.~\ref{fig:3a}, we depict the survival rate of synfire
propagation as a function of the successful transmission
probability $p$ for different values of excitatory synaptic
strength $g$, with the noise intensity $D_t=0$. We find that
each survival rate curve can be at least characterized by one
corresponding critical probability $p_{\text{on}}$. For small
$p$, due to low synaptic reliability, any synfire packet cannot
reach the final layer of the URE feedforward neuronal network.
Once the successful transmission probability $p$ exceeds the
critical probability $p_{\text{on}}$, the survival rate rapidly
transits from 0 to 1, suggesting that the propagation of
synfire activity becomes stable for a suitable high synaptic
reliability. On the other hand, besides the critical
probability $p_{\text{on}}$, we find that the survival rate
curve should be also characterized by another critical
probability $p_{\text{off}}$ if the excitatory synaptic
strength is sufficiently strong (for example, $g=3.5 $ nS in
Fig.~\ref{fig:3a}). In this case, when $p\geq p_{\text{off}}$,
our simulation results show that the survival rate rapidly
decays from 1 to 0, indicating that the network fails to
propagate the stable synfire packet again. However, it should
be noted that this does not mean that the synfire packet cannot
reach the final layer of the network in this situation, but
because the excitatory synapses with both high reliability and
strong strength lead to the occurrence of the redundant synfire
instability in transmission layers.

To systematically establish the limits for the appearance of
stable synfire propagation as well as to check that whether our
previous results can be generalized within a certain range of
parameters, we further calculate the survival rate of synfire
propagation in the $(g,p)$ panel, which is shown in
Fig.~\ref{fig:3b}. As we see, the whole $(g,p)$ panel can be
clearly divided into three regimes. These regimes include the
failed synfire propagation regime (regime I), the stable
synfire propagation regime (regime II), and the synfire
instability propagation regime (regime III). Our simulation
results reveal that transitions between these regimes are
normally very fast and therefore can be described as a sharp
transition. The data shown in Fig.~\ref{fig:3b} further
demonstrate that synfire propagation is determined by the
combination of both the successful transmission probability and
excitatory synaptic strength. For a lower synaptic reliability,
the URE feedforward neuronal network might need a larger $g$ to
support the stable propagation of synfire packet.

In reality, not only the survival rate of the synfire propagation
but also its performance is largely influenced by the successful
transmission probability and the strength of the excitatory
synapses. In Figs.~\ref{fig:3c} and \ref{fig:3d}, we present the
standard deviation of the spiking times of the output synfire
packet $\sigma_{10}$ for different values of $p$ and $g$,
respectively. Note that here we only consider parameters $p$ and
$g$ within the stable synfire propagation regime. The results
illustrated in Fig.~\ref{fig:3c} clearly demonstrate that the
propagation of synfire packet shows a better performance for a
suitable higher synaptic reliability. For the ideal case $p=1$,
the URE feedforward neuronal network even has the capability to
propagate the perfect synfire packet ($\alpha_i=100$ and
$\sigma_i=0$ ms) in the absence of noise. On the other hand, it is
also found that for a fixed $p$ the performance of synfire
propagation becomes better and better as the value of $g$ is
increased (see Fig.~\ref{fig:3d}). The above results indicate that
both high synaptic reliability and strong excitatory synaptic
strength are able to help the URE feedforward neuronal network
maintain the precision of neuronal response in the stable synfire
propagation regime.

\begin{figure}[!t]
\centering
\subfigure{\includegraphics[width=7.8cm]{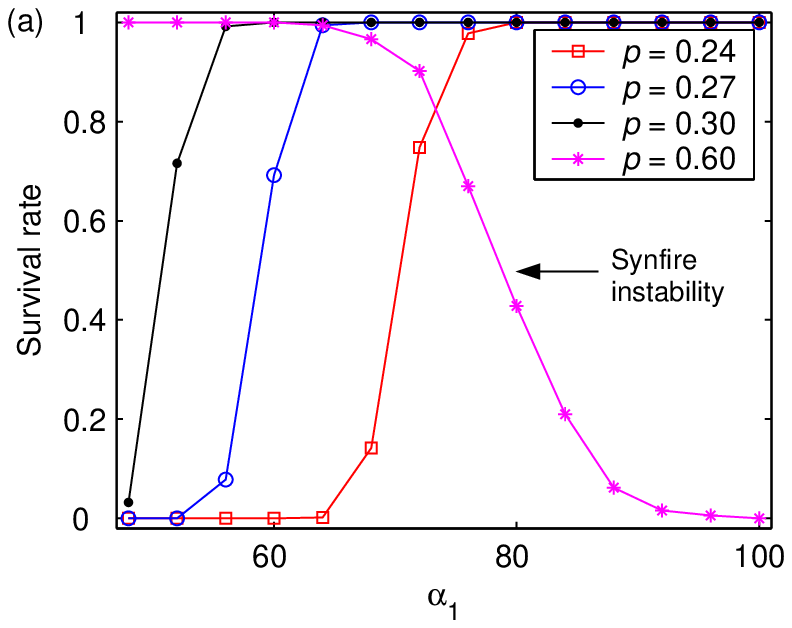}
\label{fig:4a}}
\subfigure{\includegraphics[width=7.8cm]{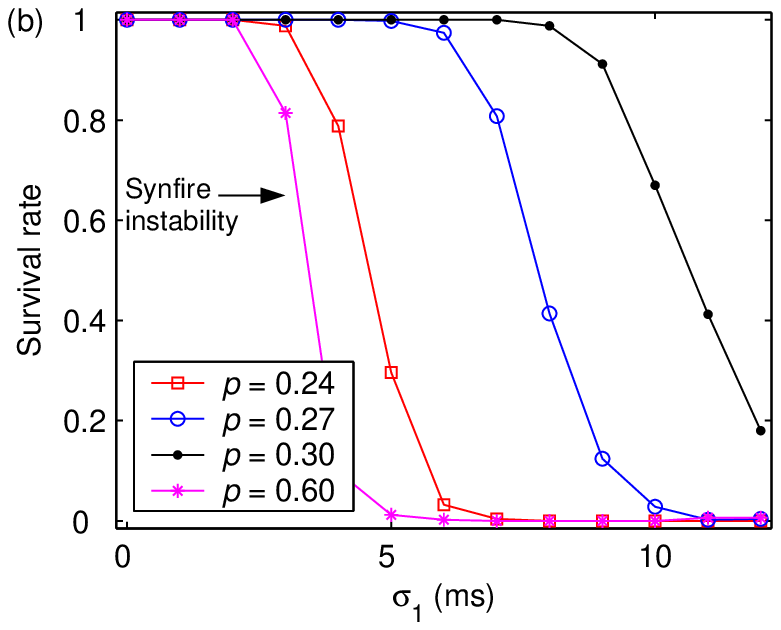}
\label{fig:4b}} \centering
\subfigure{\includegraphics[width=7.8cm]{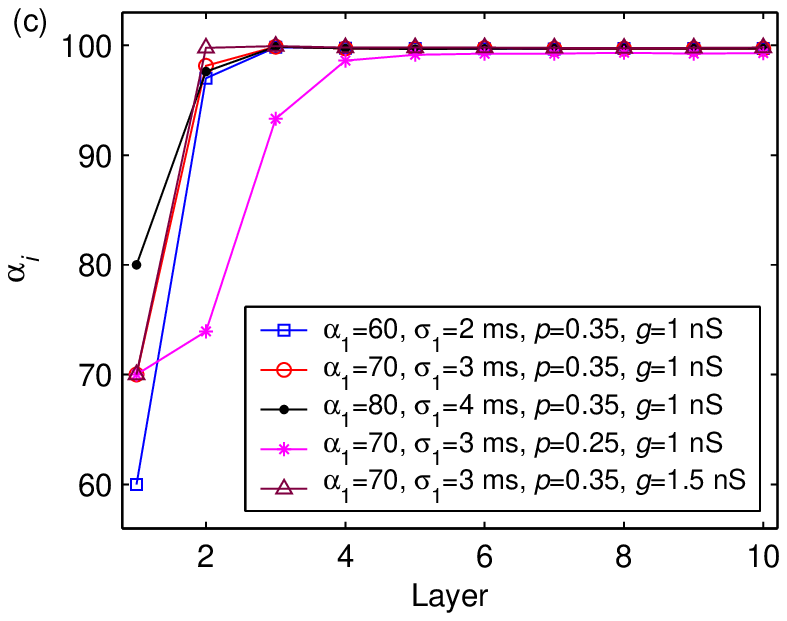}
\label{fig:4c}}
\subfigure{\includegraphics[width=7.8cm]{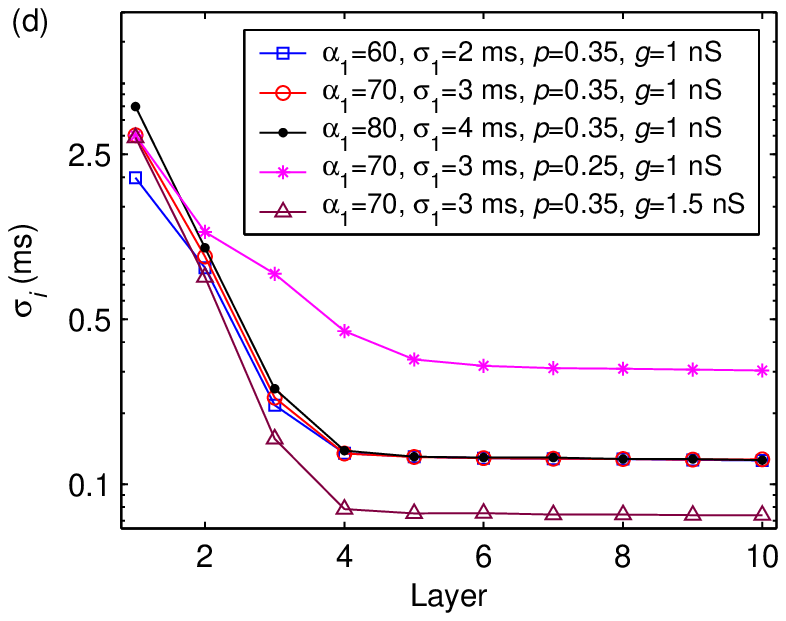}
\label{fig:4d}} \caption{\label{fig:4} (Color online)
Dependence of the synfire propagation on the parameters of the
initial spike packet. Here we display the survival rate of the
synfire propagation versus $\alpha_1$ (a) and $\sigma_1$ (b)
for different successful transmission probabilities, and the
values of $n_i$ (c) and $\sigma_i$ (d) as a function of the
layer number for different initial spike packets and different
intrinsic parameters of the network, respectively. Note that
the vertical axis in (d) is a log scale. In all cases,
the noise intensity $D_t=0$. Other parameters are $g=1$ nS and
$\sigma_1=3$ ms (a), $g=1$ nS and $\alpha_1=70$ (b). Each data
shown in (a) and (b) is computed based on 600
independent simulations, whereas each data shown in (c) and (d)
is calculated based on 500 independent successful synfire
propagation simulations.}
\end{figure}

Up to now, we only use the perfect initial spike packet
($\alpha_1=100$ and $\sigma_1=0$ ms) to evoke the synfire
propagation. This is a special case which is simplified for
analysis, but it is not necessary to restrict this condition.
To understand how a generalized spike packet is propagated
through the URE feedforward neuronal network, we randomly
choose $\alpha_1$ neurons from the sensory layer, and let each
of these neurons fire and only fire a spike at any moment
according to a Gaussian distribution with the standard
deviation $\sigma_1$. In Figs.~\ref{fig:4a} and \ref{fig:4b},
we plot the survival rate of the synfire propagation as a
function of $\alpha_1$ and $\sigma_1$ for four different values
of successful transmission probability, respectively. When the
successful transmission probability is not too large (for
example, $p=0.24$, 0.27, and 0.3 in Figs.~\ref{fig:4a}
and~\ref{fig:4b}), the synfire activity is well build up after
several initial layers for sufficiently strong initial spike
packet (large $\alpha_1$ and small $\sigma_1$), and then this
activity can be successfully transmitted along the entire
network with high survival rate. In this case, too weak initial
spike packet (small $\alpha_1$ and large $\sigma_1$) leads to
the propagation of the neural activities becoming weaker and
weaker with the increasing of layer number. Finally, the neural
activities are stopped before they reach the final layer of the
network. Moreover, with the increasing of the successful
transmission probability, neurons in the downstream layers will
share more common synaptic currents from neurons in the
corresponding upstream layers. This means that neurons in the
considered network have the tendency to fire more synchronously
for suitable larger $p$ (not too large). On the other hand, for
sufficiently high synaptic reliability (for instance, $p=0.6$
in Figs.~\ref{fig:4a} and~\ref{fig:4b}), a large $\alpha_1$ or
a suitable large $\sigma_1$ may result in the occurrence of
synfire instability, which also reduces the survival rate of
the synfire propagation. Therefore, for a fixed $g$, the URE
feedforward neuronal network with suitable higher synaptic
reliability has the ability to build up stable synfire
propagation from a slightly weaker initial spike packet (see
Figs.~\ref{fig:4a} and~\ref{fig:4b}).

Figures~\ref{fig:4c} and \ref{fig:4d} illustrate the values of
$\alpha_i$ and $\sigma_i$ versus the layer number for different
initial spike packets and several different intrinsic system
parameters of the network (the successful transmission
probability $p$ and excitatory synaptic strength $g$). For each
case shown in Figs.~\ref{fig:4c} and \ref{fig:4d}, once the
synfire propagation is successfully established, $n_i$
converges fast to the saturated value 100 and $\sigma_i$
approaches to an asymptotic value. Although the initial spike
packet indeed determines whether the synfire propagation can be
established or not as well as influences the performance of
synfire propagation in the first several layers, but it does
not determine the value of $\sigma_i$ in deep layers provided
that the synfire propagation is successfully evoked. For the
same intrinsic system parameters, if we use different initial
spike packets to evoke the synfire propagation, the value of
$\sigma_i$ in deep layers is almost the same for different
initial spike packets (see Fig.~\ref{fig:4d}). The above
results indicate that the performance of synfire propagation in
deep layers of the URE feedforward neuronal network is quite
stubborn, which is mainly determined by the intrinsic
parameters of the network but not the parameters of the initial
spike packet. In fact, many studies have revealed that the
synfire activity is governed by a stable attractor in the
$(\alpha, \sigma)$ space (Diesmann et al. 1999; Diesmann et al.
2001; Diesmann 2002; Gewaltig et al. 2001). Our above finding
is a signature that the stable attractor of synfire propagation
does also exist for the feedforward neuronal networks with
unreliable synapses.

\begin{figure}[!t]
\centering \subfigure{\includegraphics[height=6.5cm]{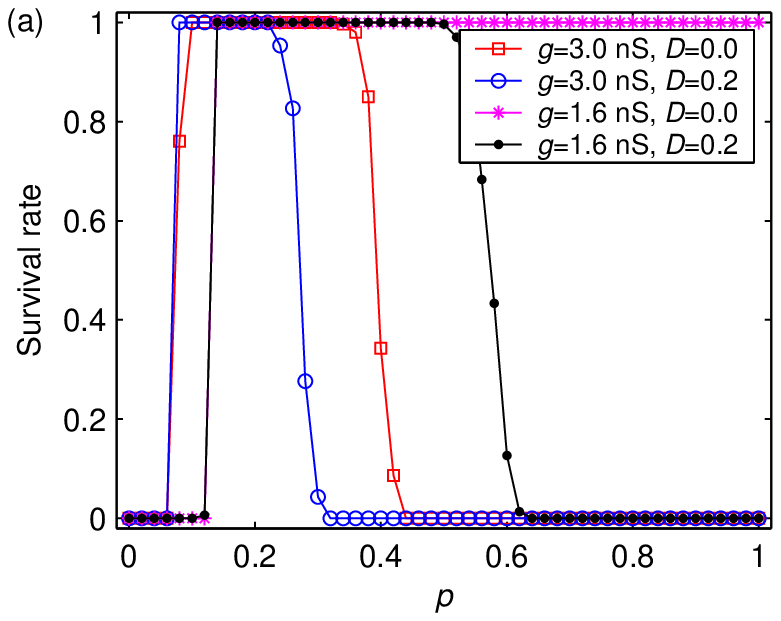} \label{fig:5a}}
\subfigure{\includegraphics[height=6.5cm]{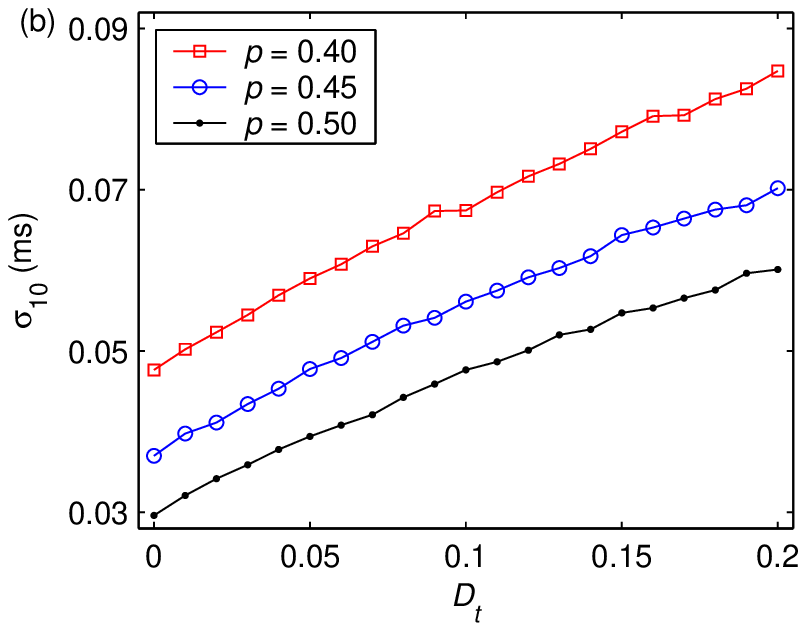} \label{fig:5b}}
\caption{\label{fig:5} (Color online) Effects of the noise intensity $D$
on the synfire propagation. (a) The survival rate versus successful
transmission probability $p$ for different excitatory coupling strength
and noise intensities. (b) The value of $\sigma_{10}$ versus the noise
intensity $D$ for different successful transmission probabilities, with
the excitatory synaptic strength $g=1.5$ nS. In all cases, the parameters
of the initial spike packet are $\alpha_1=100$ and $\sigma_1=0$ ms.
Each data shown here is computed based on 200 independent simulations
with different random seeds.}
\end{figure}

Next, we study the dependence of synfire propagation on
neuronal noise. It is found that both the survival rate of
synfire propagation and its performance are largely influenced
by the noise intensity. There is no significant qualitative
difference between the corresponding survival rate curves in
low synaptic reliable regime. However, we find important
differences between these curves for small $g$ in high synaptic
reliable regime, as well as for large $g$ in intermediate
synaptic reliable regime, that is, during the transition from
the successful synfire regime to the synfire instability
propagation regime (see from Fig.~\ref{fig:5a}). For each case,
it is obvious that the top region of the survival rate becomes
smaller with the increasing of noise intensity. This is at
least due to the following two reasons: (i) noise makes neurons
desynchronize, thus leading to a more dispersed synfire packet
in each layer. For relatively high synaptic reliability, a
dispersed synfire packet has the tendency to increase the
occurrence rate of the synfire instability. (ii) Noise with
large enough intensity results in several spontaneous neural
firing activities at random moments, which also promote the
occurrence of the synfire instability. Figure~\ref{fig:5b}
presents the value of $\sigma_{10}$ as a function of the noise
intensity $D_t$ for different values of successful transmission
probability $p$. As we see, the value of $\sigma_{10}$ becomes
larger and larger as the noise intensity is increased from 0 to
0.1 (weak noise regime). This is also due to the fact that the
existence of noise makes neurons desynchronize in each layer.
However, although noise tends to reduce the synchrony of
synfire packet, the variability of $\sigma_i$ in deep layers is
quite low (data not shown). The results suggest that, in weak
noise regime, the synfire packet can be stably transmitted
through the feedforward neuronal network with small fluctuation
in deep layers, but displays slightly worse performance
compared to the case of $D_t=0$. Further increase of noise will
cause many spontaneous neural firing activities which might
significantly deteriorate the performance of synfire
propagation. However, it should be emphasized that, although
the temporal spread of synfire packet tends to increase as the
noise intensity grows, several studies have suggested that
under certain conditions the basin of attraction of synfire
activity reaches a maximum extent (Diesmann 2002; Postma et al.
1996; Boven and Aertsen 1990). Such positive effect of noise
can be compared to a well known phenomenon called aperiodic
stochastic resonance (Collins et al. 1995b; Collins et al.
1996; Diesmann 2002).

\subsection{Firing rate propagation in URE feedforward neuronal networks}
\label{sec:3b}

In this subsection, we examine the firing rate propagation in URE
feedforward neuronal networks. To this end, we assume that all
sensory neurons are injected to a same time-varying external
current $I(t)$ (see Sec.~\ref{sec:2b} for detail). Note now that
the sensory neurons are modeled by using the integrate-and-fire
neuron model in the study of the firing rate propagation.

\begin{figure}[!t]
\centering \includegraphics[width=7.8cm]{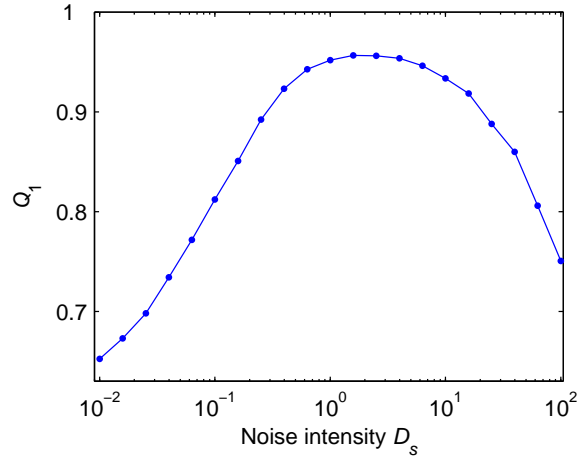}
\caption{\label{fig:6}The maximum cross-correlation coefficient
between the smooth version of external input current $I_s(t)$ and
the population firing rate of sensory neurons $r_1(t)$ for different
noise intensities.}
\end{figure}

\begin{figure}[!t]
\centering \subfigure{\includegraphics[width=6.5cm]{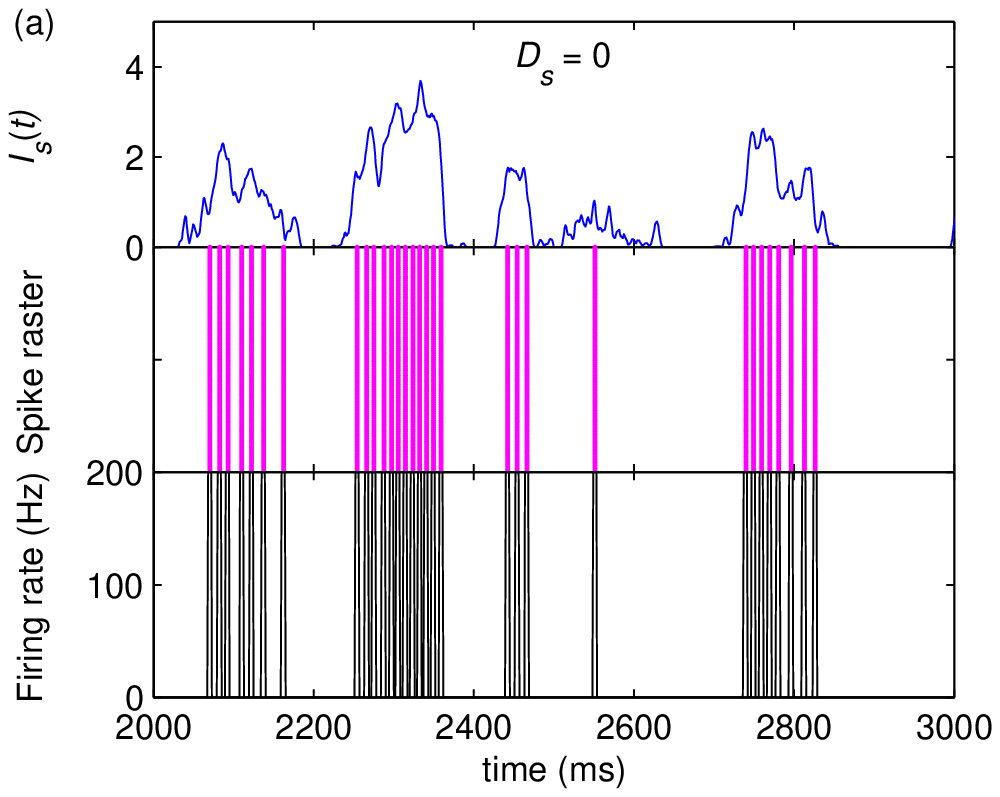}
\label{fig:7a}} \subfigure{\includegraphics[width=6.5cm]{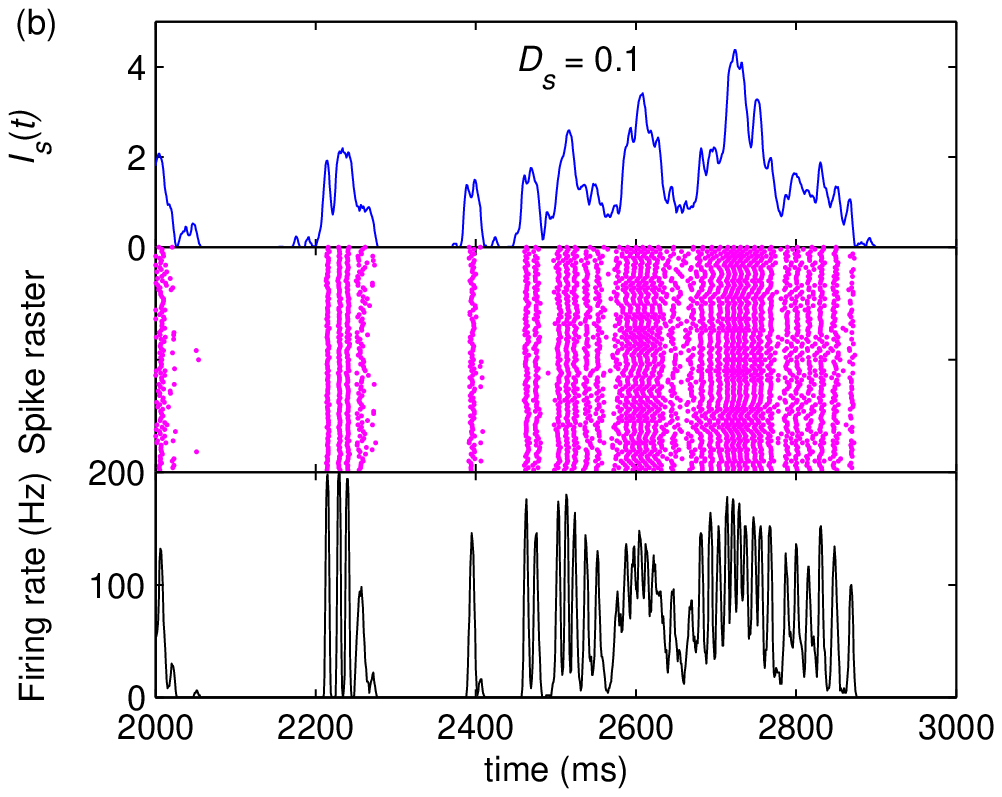}
\label{fig:7b}} \subfigure{\includegraphics[width=6.5cm]{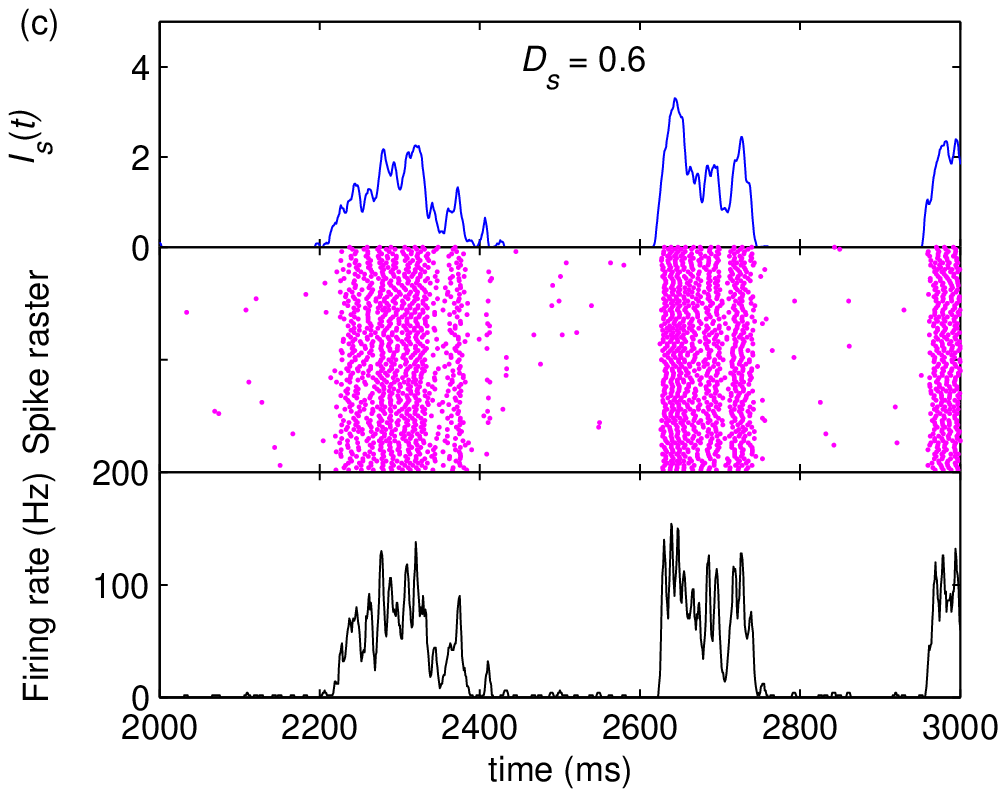}
\label{fig:7c}} \subfigure{\includegraphics[width=6.5cm]{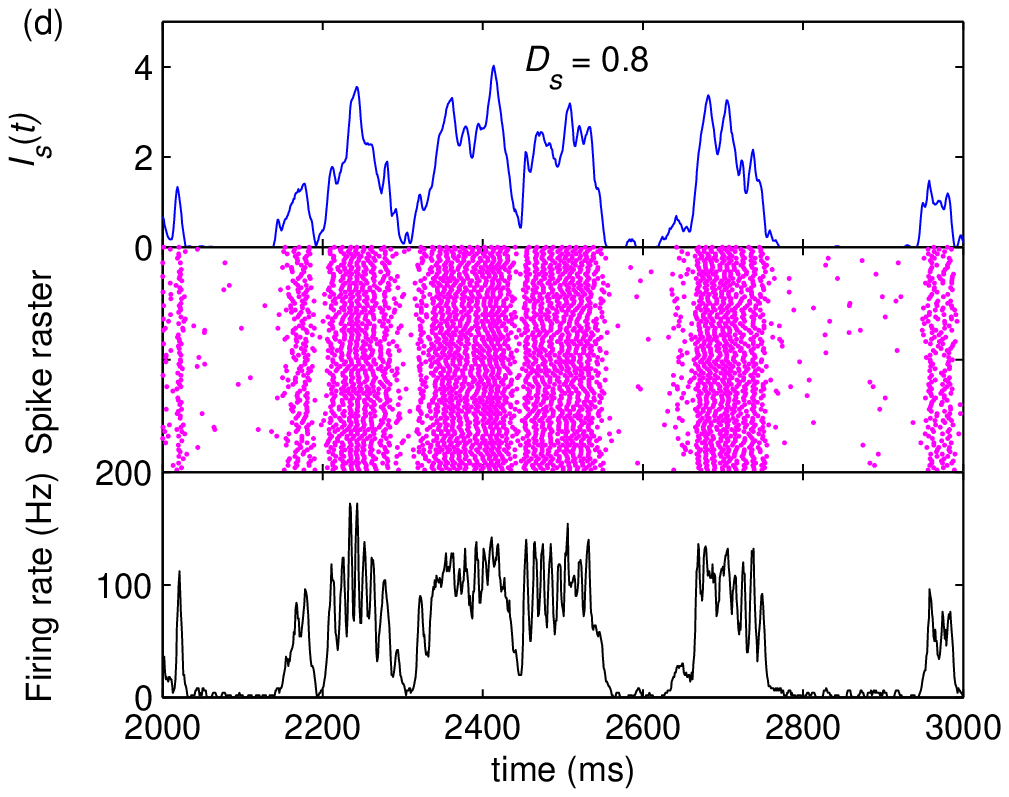}
\label{fig:7d}} \subfigure{\includegraphics[width=6.5cm]{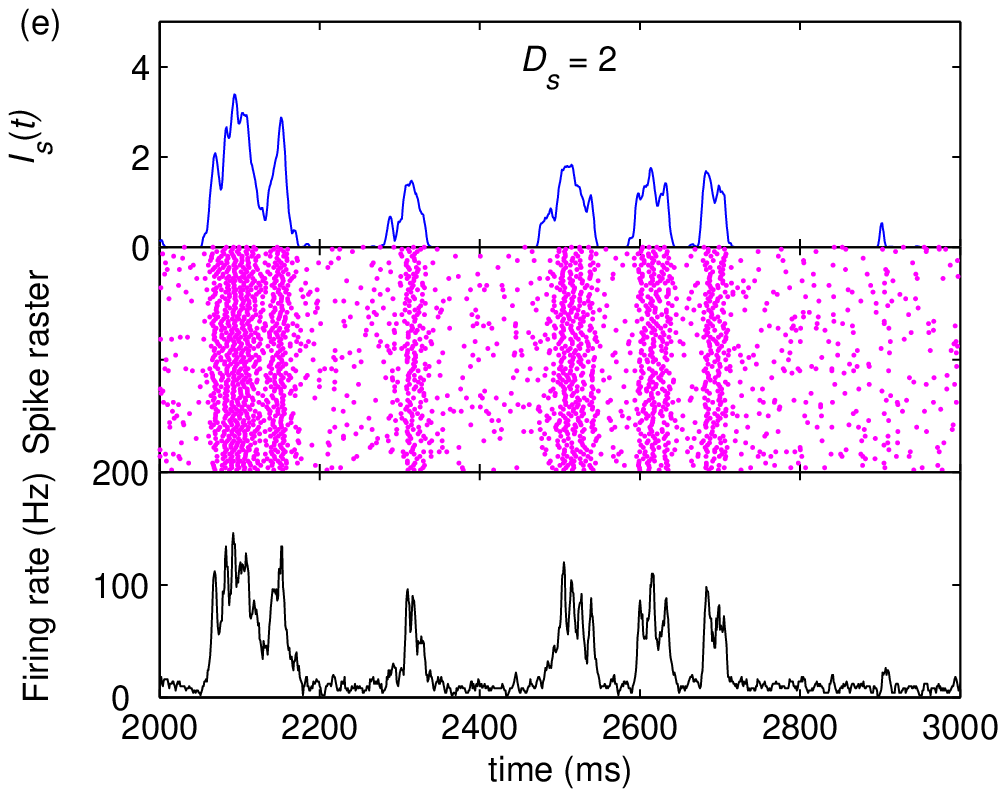}
\label{fig:7e}} \subfigure{\includegraphics[width=6.5cm]{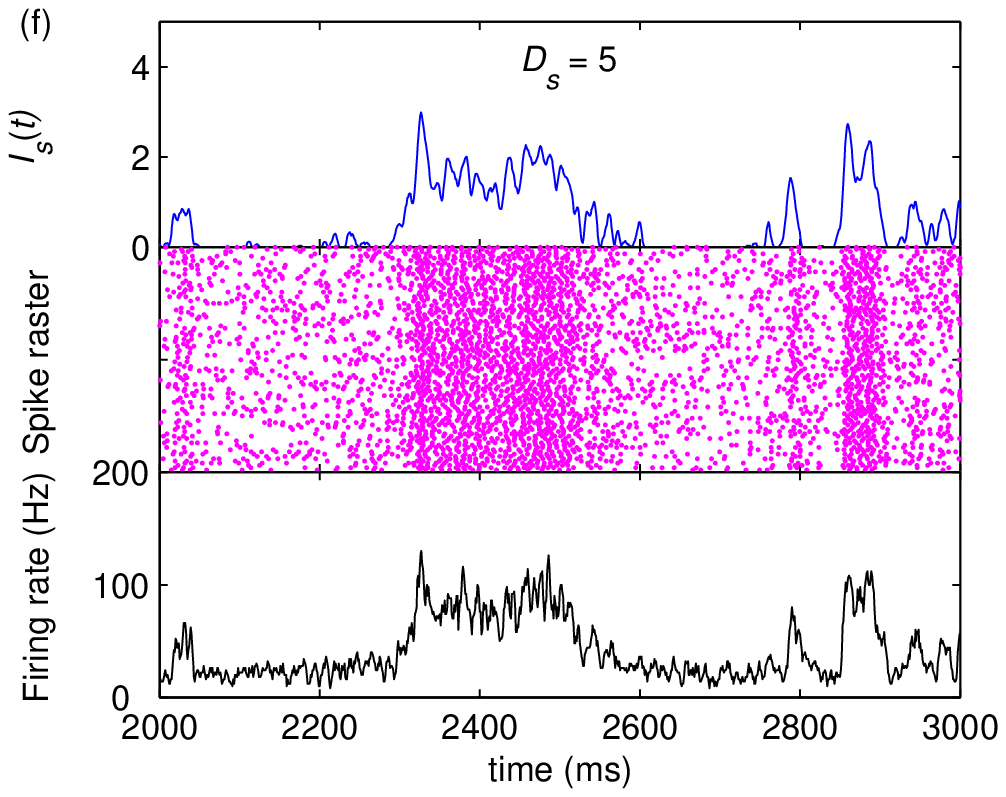}
\label{fig:7f}} \caption{\label{fig:7}(Color online) Impacts of
noise on the encoding performance (the firing rate mode) of
sensory neurons. For each case, the smooth version of the external
input current $I_s(t)$ (top panel), spike raster diagram of
sensory neurons (middle panel), and population firing rate of
sensory neurons (bottom panel) are shown. Noise intensities are
$D_s=0$ (a), $D_s=0.1$ (b), $D_s=0.6$ (c), $D_s=0.8$ (d), $D_s=2$
(e), and $D_s=5$ (f), respectively.}
\end{figure}

Before we present the results of the firing rate propagation,
let us first investigate how noise influences the encoding
capability of sensory neurons by the population firing rate.
This is an important preliminary step, because how much input
information represented by sensory neurons will directly
influence the performance of firing rate propagation. The
corresponding results are plotted in Figs.~\ref{fig:6} and
\ref{fig:7}, respectively. When the noise is too weak, the
dynamics of sensory neurons is mainly controlled by the same
external input current, which causes neurons to fire spikes
almost at the same time (see Figs.~\ref{fig:7a} and
\ref{fig:7b}). In this case, the information of the external
input current is poorly encoded by the population firing rate
since the synchronous neurons have the tendency to redundantly
encode the same aspect of the external input signal. When the
noise intensity falls within a special intermediate range
(about 0.5-10), neuronal firing is driven by both the external
input current and noise. With the help of noise, the firing
rate is able to reflect the temporal structural information
(i.e., temporal waveform) of the external input current to a
certain degree (see Figs.~\ref{fig:7c} to \ref{fig:7f}), and
therefore $Q_1$ has large value in this situation. For too
large noise intensity, the external input current is almost
drowned in noise, thus resulting that the input information
cannot be well read from the population firing rate of sensory
neurons again. On the other hand, sensory neurons can fire
``false'' spikes provided that they are driven by sufficiently
strong noise (as for example at $t\approx2800$ ms in
Fig.~\ref{fig:7e}). Although the encoding performance of the
sensory neurons might be good enough in this case, our
numerical simulations reveal that such false spikes will
seriously reduce the performance of the firing rate propagation
in deep layers, which will be discussed in detail in the later
part of this section. By taking these factors into account, we
consider the noise intensity of sensory neurons to be within
the range of 0.5 to 1 in the present work.

\begin{figure}[!t]
\centering \includegraphics[scale=0.78]{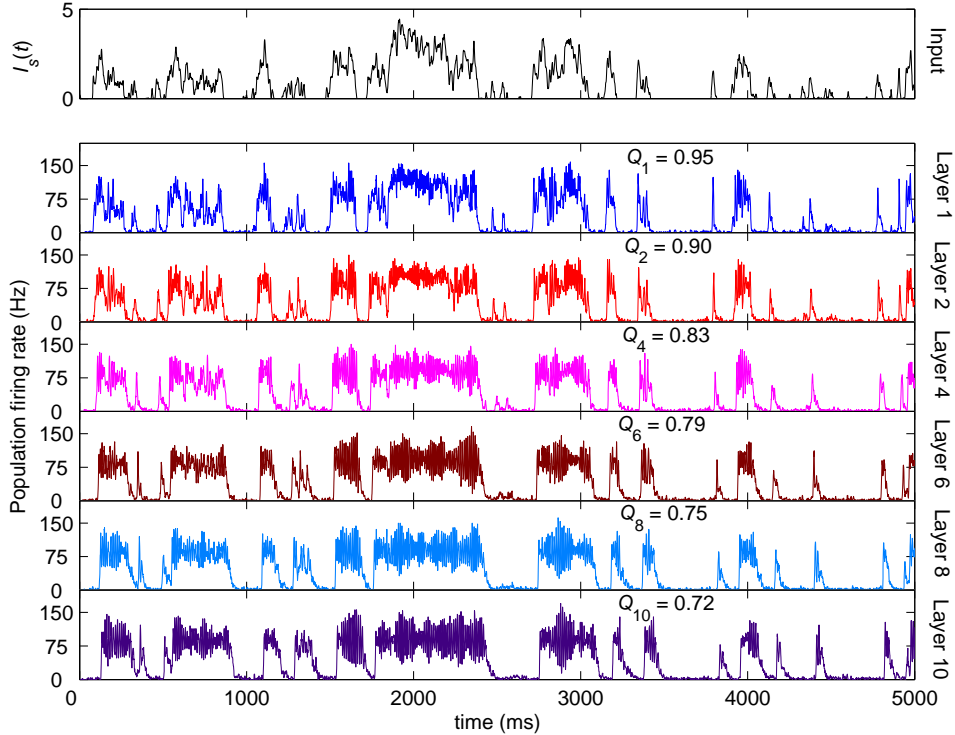}
\caption{\label{fig:8}  An example of the firing rate propagation
in the URE feedforward network. Here we show the smooth version of
the external input current $I_s(t)$, as and the population firing
rates of layers 1, 2, 4, 6, 8, and 10, respectively. System
parameters are $g=0.4$ nS, $p=0.2$, and $D_s=D_t=0.7$.}
\end{figure}

Figure~\ref{fig:8} shows a typical example of the firing rate
propagation. In view of the overall situation, the firing rate can
be propagated rapidly and basically linearly in the URE
feedforward neuronal network. However, it should be noted that,
although the firing rates of neurons from the downstream layers
tend to track those from the upstream layers, there are still
several differences between the firing rates for neurons in two
adjacent layers. For example, it is obvious that some low firing
rates may disappear or be slightly amplified in the first several
layers, as well as some high firing rates are weakened to a
certain degree during the whole transmission process. Therefore,
as the neural activities are propagated across the network, the
firing rate has the tendency to lose a part of local detailed
neural information but can maintain a certain amount of global
neural information. As a result, the maximum cross-correlation
coefficient between $I_s(t)$ and $r_i(t)$ basically drops with the
increasing of the layer number.

\begin{figure}[!t]
\centering \subfigure{\includegraphics[width=7.8cm]{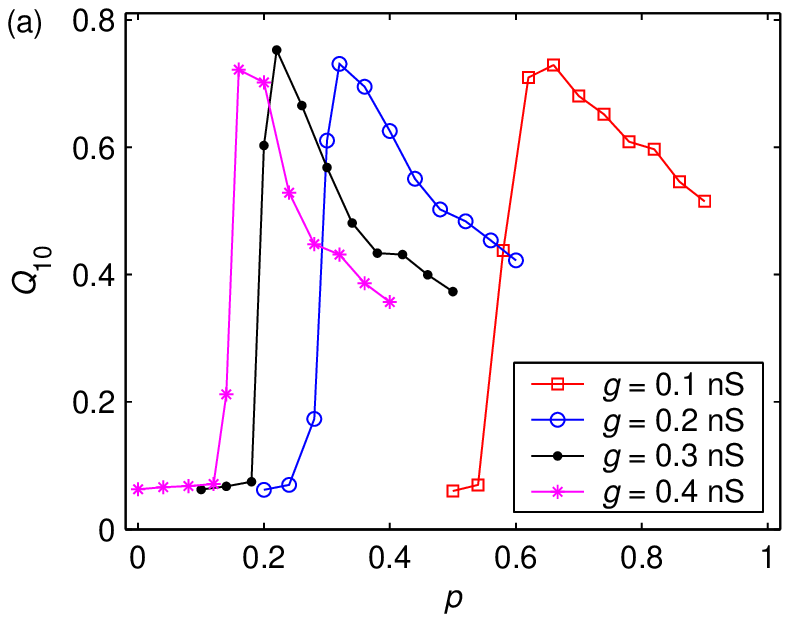} \label{fig:9a}}
\subfigure{\includegraphics[width=7.8cm]{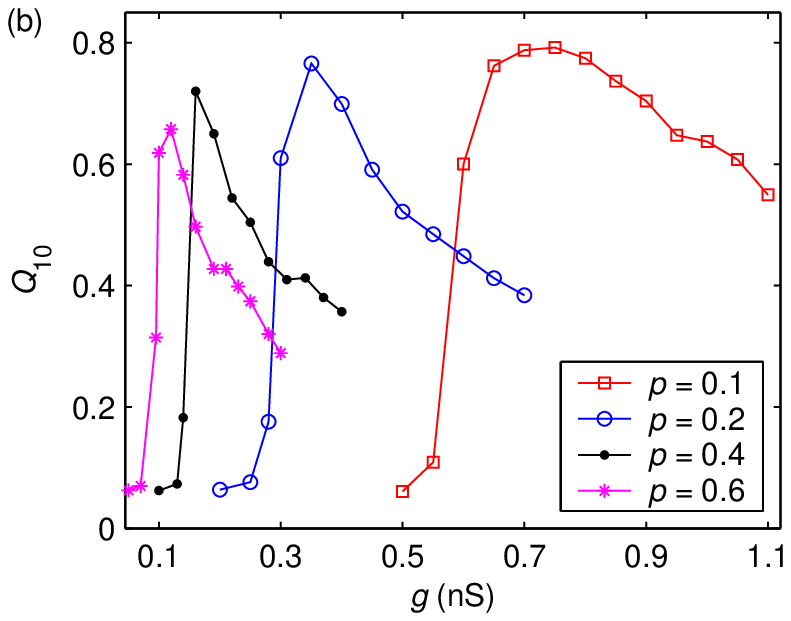} \label{fig:9b}}
\caption{\label{fig:9} (Color online) Effects of the unreliable
synapses on the performance of firing rate propagation. (a) The
value of $Q_{10}$ as a function of $p$ for different values of
excitatory synaptic strength. (b) The value of $Q_{10}$ as a
function of $g$ for different values of successful transmission
probability. Noise intensities are $D_s=D_t=0.5$ in all cases.
Here each data point is computed based on 50 different independent
simulations with different random seeds.}
\end{figure}

Let us now assess the impacts of the unreliable synapses on the
performance of firing rate propagation in the URE feedforward
neuronal network. Figure~\ref{fig:9a} presents the value of
$Q_{10}$ versus the success transmission probability $p$ for
various excitatory synaptic strengths. For a fixed value of $g$, a
bell-shaped $Q_{10}$ curve is clearly seen by changing the value
of successful transmission probability, indicating that the firing
rate propagation shows the best performance at an optimal synaptic
reliability level. This is because, for each value of $g$, a very
small $p$ will result in the insufficient firing rate propagation
due to low synaptic reliability, whereas a sufficiently large $p$
can lead to the excessive propagation of firing rate caused by
burst firings. Based on above reasons, the firing rate can be well
transmitted to the final layer of the URE feedforward neuronal
network only for suitable intermediate successful transmission
probabilities. Moreover, with the increasing of $g$, the
considered network needs a relatively small $p$ to support the
optimal firing rate propagation. In Fig.~\ref{fig:9b}, we plot the
value of $Q_{10}$ as a function of the excitatory synaptic
strength $g$ for different values of $p$. Here the similar results
as those shown in Fig.~\ref{fig:9a} can be observed. This is due
to the fact that increasing $g$ and fixing the value of $p$ is
equivalent to increasing $p$ and fixing the value of $g$ to a
certain degree. According to the aforementioned results, we
conclude that both the successful transmission probability and
excitatory synaptic strength are critical for firing rate
propagation in URE feedforward networks, and better choosing of
these two unreliable synaptic parameters can help the cortical
neurons encode neural information more accurately.

\begin{figure}[!t]
\centering \subfigure{\includegraphics[width=7.8cm]{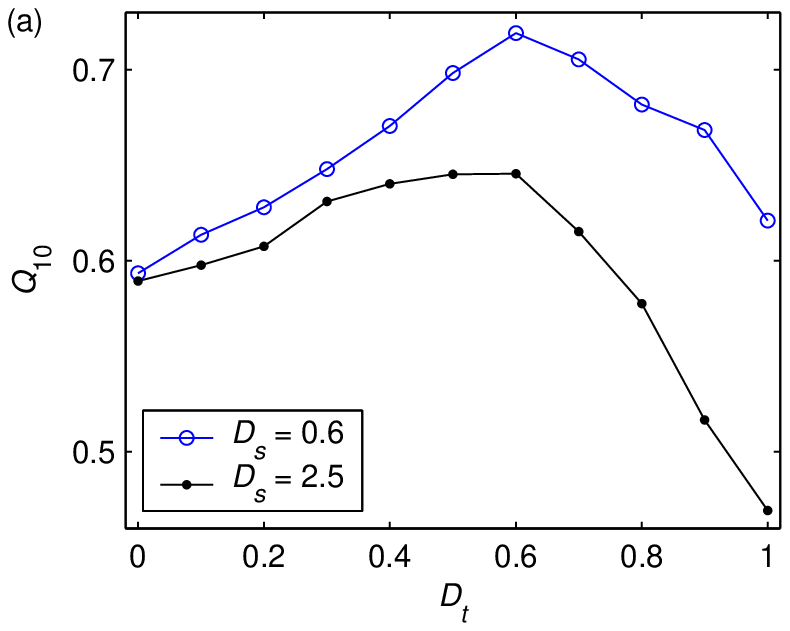} \label{fig:10a}}
\subfigure{\includegraphics[width=7.8cm]{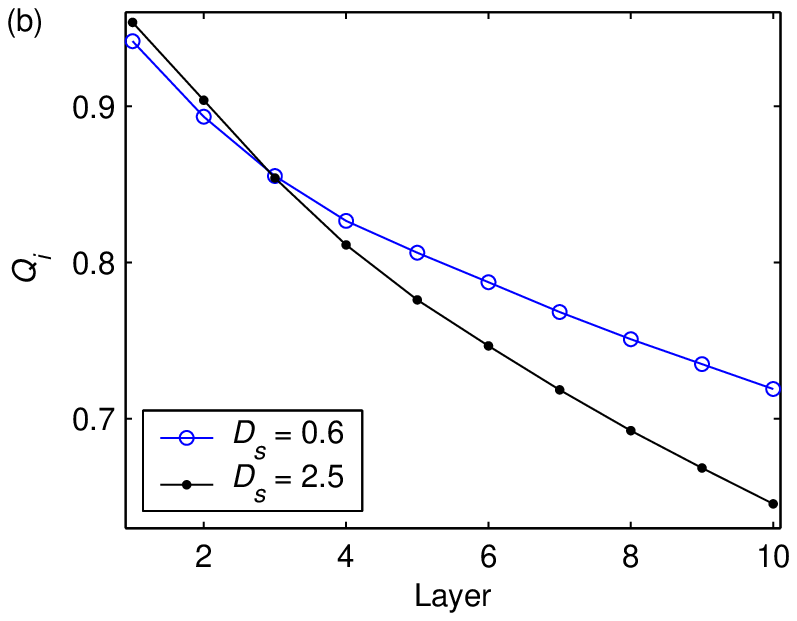} \label{fig:10b}}
\caption{\label{fig:10} (Color online) Impacts of noise
on the performance of firing rate propagation. (a) The value of
$Q_{10}$ as a function of the noise intensity of cortical neurons
$D_t$ for different values of $D_s$. (b) The performance of firing
rate propagation in each layer at $D_t=0.6$ for two different noise
intensities of sensory neurons. In all cases, $p=0.2$ and $g=0.4$ nS.
Here each data point is computed based on 50 different
independent simulations with different random seeds.}
\end{figure}

\begin{figure}[!t]
\centering \includegraphics[width=12cm]{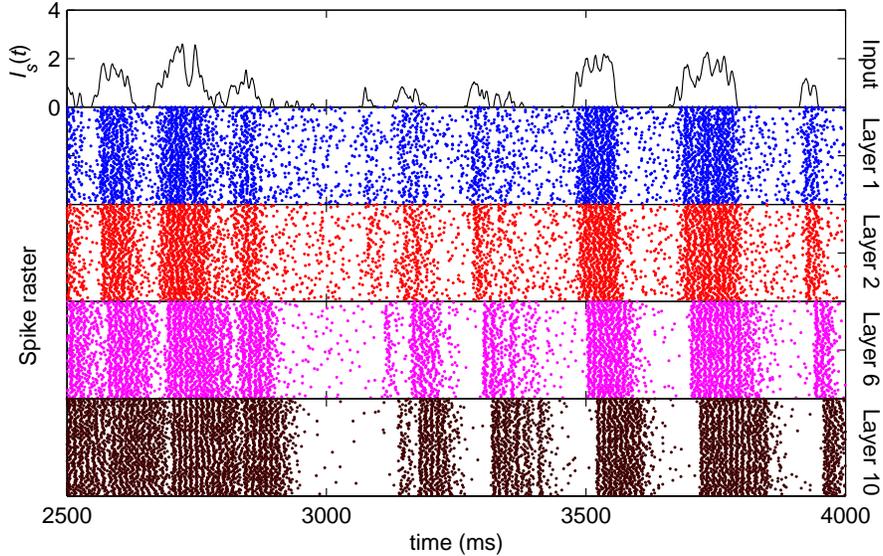}
\caption{\label{fig:11} (Color online) An example of weak
external input signal amplification. System parameters are the
successful transmission probability $p=0.2$, excitatory
synaptic strength $g=0.4$ nS, and noise intensities $D_s=2.5$
and $D_t=0.6$, respectively.}
\end{figure}

Next, we examine the dependence of the firing rate propagation on
neuronal noise. The corresponding results are plotted in
Figs.~\ref{fig:10} and \ref{fig:11}, respectively.
Figure~\ref{fig:10a} demonstrates that the noise of cortical
neurons plays an important role in firing rate propagation. Noise
of cortical neurons with appropriate intensity is able to enhance
their encoding accuracy. It is because appropriate intermediate
noise, on the one hand, prohibits synchronous firings of cortical
neurons in deep layers, and on the other hand, ensures that the
useful neural information does not drown in noise. However, the
level of enhancement is largely influenced by the noise intensity
of sensory neurons. As we see, for a large value of $D_s$, such
enhancement is weakened to a great extent. This is because
slightly strong noise intensity of sensory neurons will cause
these neurons to fire several false spikes and a part of these
spikes can be propagated to the transmission layers. If enough
false spikes appear around the weak components of the external
input current, these spikes will help the network abnormally
amplify these weak components during the whole transmission
process. The aforementioned process can be seen clearly from an
example shown in Fig.~\ref{fig:11}. As a result, the performance
of the firing rate propagation might be seriously deteriorated in
deep layers. However, it should be noted that this kind of
influence typically needs the accumulation of several layers. Our
simulation results show that the performance of firing rate
propagation can be well maintained or even becomes slightly better
(depending on the noise intensity of sensory neurons, see
Fig.~\ref{fig:6}) in the first several layers for large $D_s$ (see
Fig.~\ref{fig:10b}). In fact, the above results are based on the
assumption that each cortical neuron is driven by independent
noise current with the same intensity. Our results can be
generalized from the sensory layer to the transmission layers if
we suppose that noise intensities for neurons in different
transmission layers are different. All these results imply that
better tuning of the noise intensities of both the sensory and
cortical neurons can enhance the performance of firing rate
propagation in the URE feedforward neuronal network.

\subsection{Stochastic effect of neurotransmitter release} \label{sec:3c}

\begin{figure}[!t]
\centering \subfigure{\includegraphics[width=5.81cm]{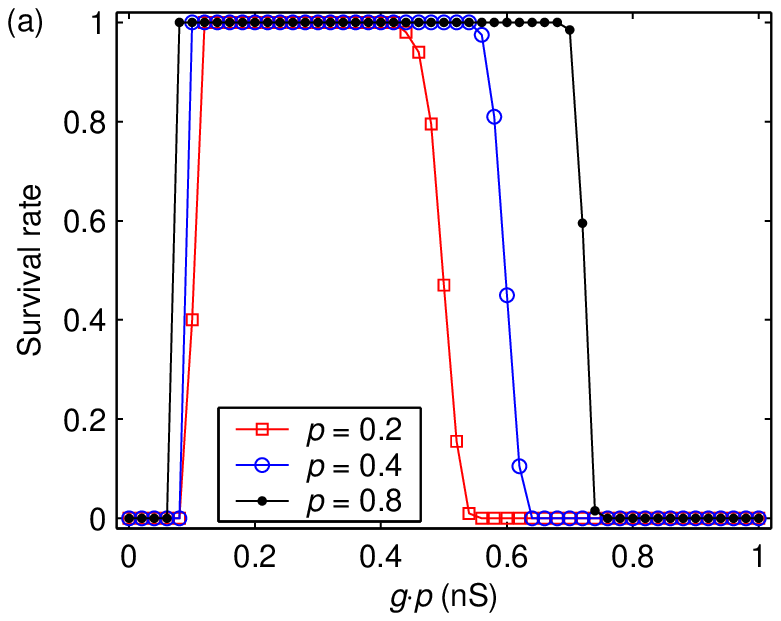}
\label{fig:12a}}
\subfigure{\includegraphics[width=5.81cm]{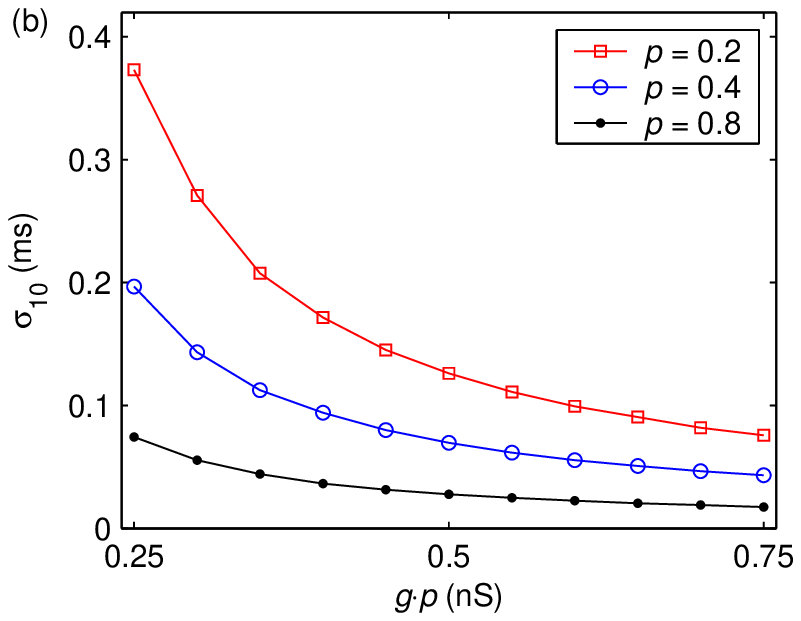}
\label{fig:12b}}
\subfigure{\includegraphics[width=5.81cm]{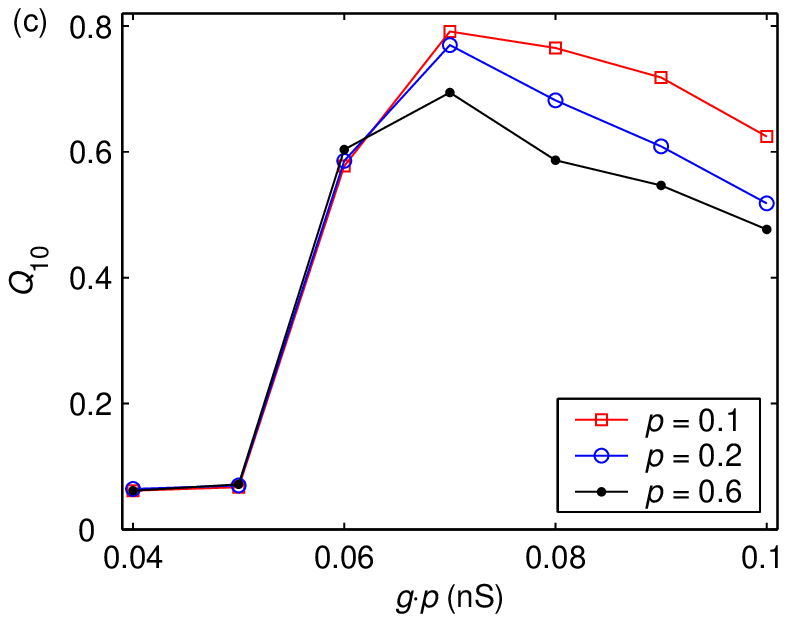}
\label{fig:12c}} \caption{\label{fig:12} (Color online)
Dependence of signal propagation dynamics on the product of $g
\cdot p$ in the URE feedforward neuronal network.
\emph{Synfire} mode: survival rate (a) and $\sigma_{10}$ versus
$g \cdot p$ with $D_t=0$. The parameters of the initial spike
packet are $\alpha_1=100$ and $\sigma_1=0$ ms. \emph{Firing
rate} mode: $Q_{10}$ as a function of $g \cdot p$ with
$D_s=D_t=0.5$. Here each data point shown in (a) and (b) is
calculated based on 200 different independent simulations,
whereas each data point shown in (c) are based on 50 different
independent simulations.}
\end{figure}

From the numerical results depicted in Secs.~\ref{sec:3a} and
\ref{sec:3b}, we find that increasing $g$ with $p$ fixed has
similar effects as increasing $p$ while keeping $g$ fixed for both
the synfire mode and firing rate mode. Some persons might
therefore postulate that the signal propagation dynamics in
feedforward neuronal networks with unreliable synapses can be
simply determined by the average amount of received
neurotransmitter for each neuron in a time instant, which can be
reflected by the product of $g\cdot p$. To check whether this is
true, we calculate the measures of these two signal propagation
modes as a function of $g \cdot p$ for different successful
transmission probabilities. If this postulate is true, the URE
feedforward neuronal network will show the same propagation
performance for different values of $p$ at a fixed $g\cdot p$. Our
results shown in Figs.~\ref{fig:12a}-\ref{fig:12c} clearly
demonstrate that the signal propagation dynamics in the considered
network can not be simply determined by the product $g\cdot p$ or,
equivalently, by the average amount of received neurotransmitter
for each neuron in a time instant. For both the synfire
propagation and firing rate propagation, although the propagation
performance exhibits the similar trend with the increasing of
$g\cdot p$, the corresponding measure curves do not superpose in
most parameter region for each case, and in some parameter region
the differences are somewhat significant (see Figs.~\ref{fig:12b}
and \ref{fig:12c}). This is because of the stochastic effect of
neurotransmitter release, that is, the unreliability of
neurotransmitter release will add randomness to the system.
Different successful transmission probabilities may introduce
different levels of randomness, which will further affect the
nonlinear spiking dynamics of neurons. Therefore, the URE
feedforward neuronal network might display different propagation
performance for different values of $p$ even at a fixed $g\cdot
p$. If we set the value of $g\cdot p$ constant, a low synaptic
reliability will introduce large fluctuations in the synaptic
inputs. For small $p$, according to the above reason, some neurons
will fire spikes more than once in the large $g\cdot p$ regime.
This mechanism increases the occurrence rate of the synfire
instability. Thus, the URE feedforward neuronal network has the
tendency to stop the stable synfire propagation for a small
synaptic transmission probability (see Fig.~\ref{fig:12a}). On the
other hand, a high synaptic reliability will introduce small
fluctuations in the synaptic inputs for a fixed $g\cdot p$. This
makes neurons in the considered network fire spikes almost
synchronously for a large $p$, thus resulting the worse
performance for the firing rate propagation in large $g\cdot p$
regime (see Fig.~\ref{fig:12c}). Our above results suggest that
the performance of the signal propagation in feedforward neuronal
networks with unreliable synapses is not only purely determined by
the change of synaptic parameters, but also largely influenced by
the stochastic effect of neurotransmitter release.

\subsection{Comparison with corresponding RRE feedforward neuronal networks}
\label{sec:3d}

In this subsection, we make comparisons on the propagation
dynamics between the URE and the RRE feedforward networks. We
first introduce how to generate a corresponding RRE feedforward
neuronal network for a given URE feedforward neuronal network.
Suppose now that there is a URE feedforward neuronal network
with successful transmission probability $p$. A corresponding
RRE feedforward neuronal network is constructed by using the
connection density $p$ (on the whole), that is, a synapse from
one neuron in the upstream layer to one neuron in the
corresponding downstream layer exists with probability $p$. As
in the URE feedforward neuronal network given in
Sec.~\ref{sec:2a}, there is no feedback connection from
downstream neurons to upstream neurons and also no connection
among neurons within the same layer in the RRE feedforward
neuronal network. It is obvious that parameter $p$ has
different meanings in these two different feedforward neuronal
network models. The synaptic interactions between neurons in
the RRE feedforward neuronal network are also implemented by
using the conductance-based model (see Eqs.~(\ref{eq:6}) and
(\ref{eq:7}) for detail). However, here we remove the
constraint of the synaptic reliability parameter for the RRE
feedforward neuronal network, e.g., $h(i,j;k,j-1)=1$ in all
cases. A naturally arising question is what are the
differences, if have, between the synfire propagation and
firing propagation in URE feedforward neuronal networks and
those in RRE feedforward neuronal networks, although the
numbers of active synaptic connections that taking part in
transmitting spikes in a time instant are the same from the
viewpoint of mathematical expectation.

\begin{figure}[!t]
\centering \includegraphics[width=7.8cm]{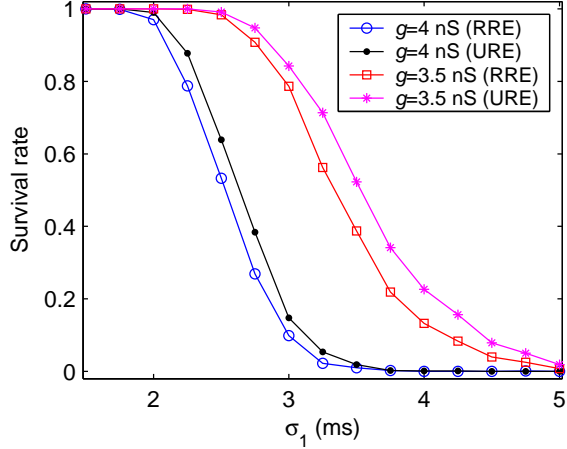}
\caption{\label{fig:13} (Color online) The difference between the
synfire propagation in the URE feedforward neuronal network and
the RRE feedforward neuronal network. Here we show the value of
survival rate as a function of $\sigma_1$ for different network
models. In all cases, $D_t=0$ and $\alpha_1=70$. Other system
parameters are $g=4$ nS and $p=0.15$ (dot: ``$\bullet$'', and
circle: ``$\circ$'' ), and $g=3.5$ nS and $p=0.12$ (square:
``$\square$'', and asterisk: ``$\ast$''). Each data point is
calculated based on 500 different independent simulations with
different random seeds.}
\end{figure}

For the synfire propagation, our simulation results indicate
that, compared to the RRE feedforward neuronal network, the URE
feedforward neuronal network is able to suppress the occurrence
of synfire instability to a certain degree, which can be seen
clearly in Fig.~\ref{fig:13}. Typically, this phenomenon can be
observed in strong excitatory synaptic strength regime. Due to
the heterogeneity of connectivity, some neurons in the RRE
feedforward neuronal network will have more input synaptic
connections than the other neurons in the same network. For
large value of $g$, these neurons tend to fire spikes very
rapidly after they received synaptic currents. If the width of
the initial spike packet is large enough, these neurons might
fire spikes again after their refractory periods, which are
induced by a few spikes from the posterior part of the
dispersed initial spike packet. These spikes may increase the
occurrence rate of the synfire instability. While in the case
of URE feedforward neuronal network, the averaging effect of
unreliable synapses tends to prohibit neurons fire spikes too
quickly. Therefore, under the equivalent parameter conditions,
less neurons can fire two or more spikes in the URE feedforeard
neuronal network. As a result, the survival rate of the synfire
propagation for the URE feedforeard neuronal network is larger
than that for the RRE feedforward neuronal network (see
Fig.~\ref{fig:13}), though not so significant.

\begin{figure}[!t]
\centering \subfigure{\includegraphics[width=7.8cm]{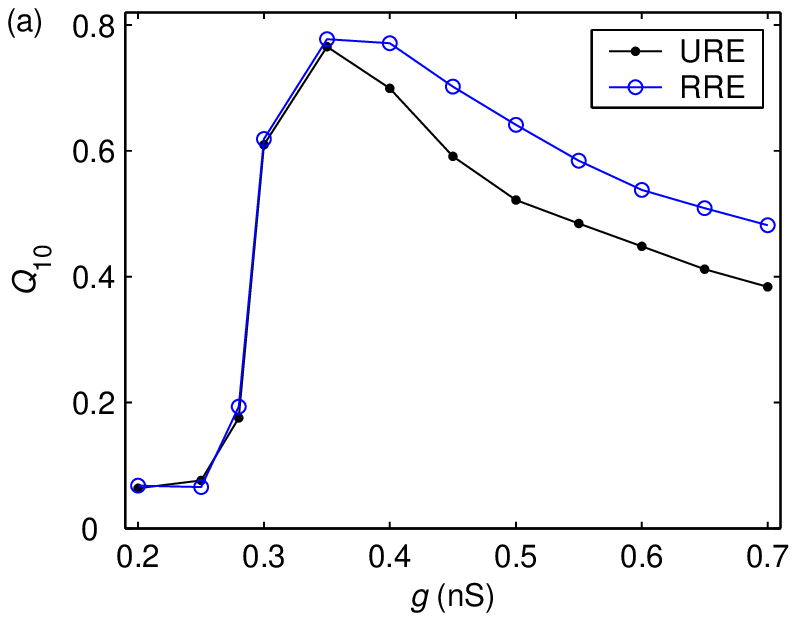}
\label{fig:14a}} \subfigure{\includegraphics[width=7.8cm]{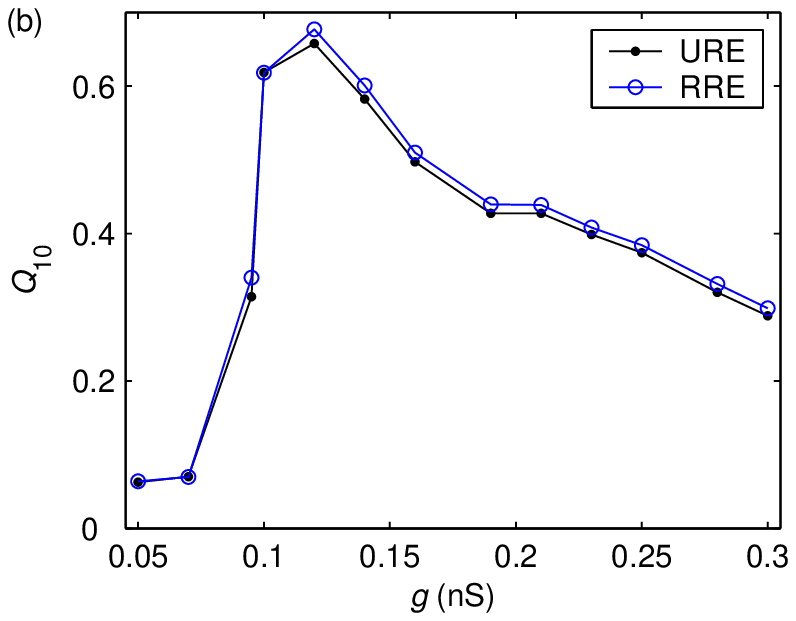}
\label{fig:14b}} \caption{\label{fig:14} (Color online) Firing
rate propagation in URE feedforward neuronal network and RRE
feedforward neuronal network. The value of $Q_{10}$ as a function
of excitatory synaptic strength $g$ for $p=0.2$ (a) and $p=0.6$
(b), respectively. Noise intensities are $D_t=D_s=0.5$. Each data
point is calculated based on 50 different independent simulations
with different random seeds.}
\end{figure}

In further simulations, we find interesting results in small
$p$ regime for the firing rate propagation. Compared to the
case of the URE feedforward neuronal network, the RRE
feedforward neuronal network can better support the firing rate
propagation in this small $p$ regime for strong excitatory
synaptic strength (see Fig.~\ref{fig:14a}). It is because the
long-time averaging effect of unreliable synapses at small $p$
tends to make neurons fire more synchronous spikes in the URE
feedforward neuronal network through the homogenization process
of synaptic currents. However, with the increasing of $p$,
neurons in the downstream layers have the tendency to share
more common synaptic currents from neurons in the corresponding
upstream layers for both types of feedforward neuronal
networks. The aforementioned factor makes the difference of the
performance of firing rate propagation between these two types
of feedforward neuronal networks become small so that the
$Q_{10}$ curves almost coincide with each other for the case of
$p=0.6$ (see Fig.~\ref{fig:14b}).

Although from the above results we can not conclude that
unreliable synapses have advantages and play specific
functional roles in signal propagation, not like those results
shown in the previous studies (Goldman et al. 2002; Goldman
2004), at least it is shown that the signal propagation
activities are different in URE and RRE to certain degrees. We
should be cautioned when using random connections to replace
unreliable synapses in modelling research. However, it should
be noted that the RRE feedforward neuronal network considered
here is just one type of diluted feedforward neuronal networks.
There exists several other possibilities to construct the
corresponding diluted feedforward neuronal networks (Hehl et
al. 2001). The similar treatments for these types of diluted
feedforward neuronal networks require further investigation.

\subsection{Signal propagation in UREI feedforward neuronal networks}
\label{sec:3e}

In this subsection, we further study the signal propagation in
the feedforward neuronal networks composed of both excitatory
and inhibitory neurons connected in an all-to-all coupling
fashion (i.e., the UREI feedforward neuronal networks). This
study is necessary because real biological neuronal networks,
especially mammalian neocortex, consist not only of excitatory
neurons but also of inhibitory neurons. The UREI feedforward
neuronal network studied in this subsection has the same
topology as that shown in Fig.~\ref{fig:1}. In simulations, we
randomly choose 80 neurons in each layer as excitatory and the
rest of them as inhibitory, as the ratio of excitatory to
inhibitory neurons is about $4:1$ in mammalian neocortex. The
dynamics of the unreliable inhibitory synapse is also modeled
by using Eqs.~(\ref{eq:6}) and (\ref{eq:7}). The reversal
potential of the inhibitory synapse is fixed at -75 mV, and its
strength is set as $J=K \cdot g$, where $K$ is a scale factor
used to control the relative strength of inhibitory and
excitatory synapses. Since the results of the signal
propagation in UREI feedforward neuronal networks are quite
similar to those in URE feedforward neuronal networks, we omit
most of them and only discuss the effects of inhibition in
detail.

\begin{figure}[!t]
\centering \subfigure{\includegraphics[width=6.5cm]{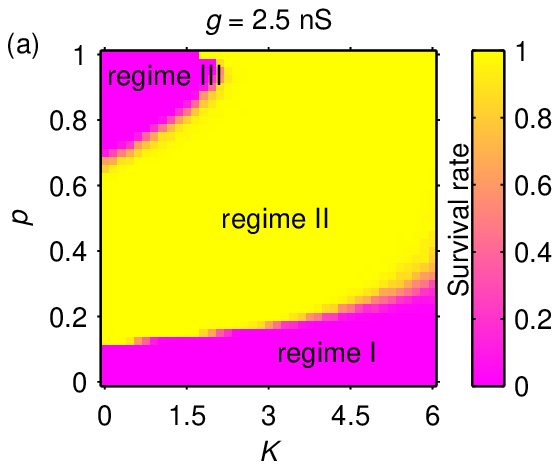}
\label{fig:15a}} \subfigure{\includegraphics[width=6.5cm]{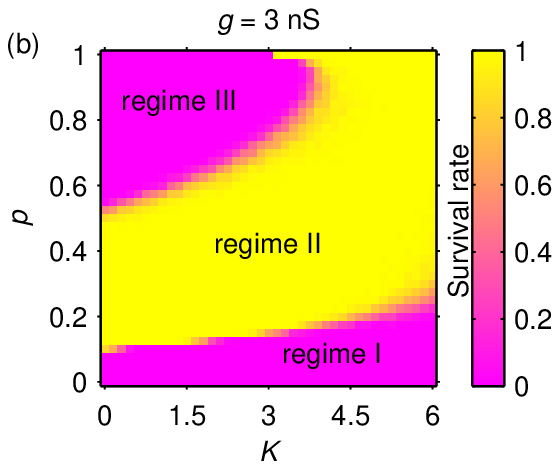}
\label{fig:15b}} \subfigure{\includegraphics[width=6.5cm]{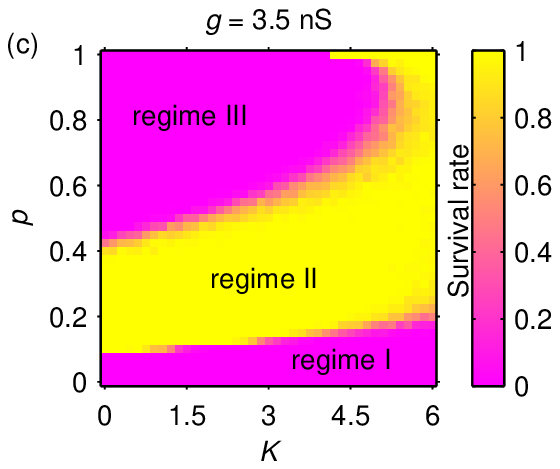}
\label{fig:15c}} \caption{\label{fig:15} (Color online) Partition
of three different synfire propagation regimes in the $(K,p)$
panel ($41\times41=1681$ points). Regime I: the failed synfire
propagation region; regime II: the stable synfire propagation
region; and regime III: the synfire instability propagation region.
we set $g=2.5$ nS (a), $g=3$ nS (b), and $g=3.5$ nS (c). In all
cases, the parameters of the initial spike packet are
$\alpha_1=100$ and $\sigma_1=0$ ms, and the noise intensity is
$D_t=0$. Each data point shown here is calculated based on 200
different independent simulations with different random seeds.}
\end{figure}

Figure~\ref{fig:15} shows the survival rate of synfire
propagation in the $(K,p)$ panel for three different excitatory
synaptic strengths. Depending on whether the synfire packet can
be successfully and stably transmitted to the final layer of
the UREI feedforward neuronal network, the whole $(K,p)$ panel
can also be divided into three regimes. For each considered
case, the network with both small successful transmission
probability and strong relative strength of inhibitory and
excitatory synapses (failed synfire regime) prohibits the
stable propagation of the synfire activity. While in the case
of high synaptic reliability and small $K$ (synfire instability
propagation regime), the synfire packet also cannot be stably
transmitted across the whole network due to the occurrence of
synfire instability. Therefore, the UREI feedforward neuronal
network is able to propagate the synfire activity successfully
in a stable way only for suitable combination of parameters $p$
and $K$. Moreover, due to the competition between excitation
and inhibition, the transitions between these different regimes
cannot be described as a sharp transition anymore, in
particular, for large scale factor $K$. Our results suggest
that such non-sharp character is strengthen with the increasing
of $g$. On the other hand, the partition of these different
propagation regimes depends not only on parameters $p$ and $K$
but also on the excitatory synaptic strength $g$. As the value
of $g$ is decreased, both the synfire instability propagation
regime and stable synfire propagation regime are shifted to the
upper left of the $(K,p)$ panel at first, and then disappear
one by one (data not shown). In contrast, a strong excitatory
synaptic strength has the tendency to extend the areas of the
synfire instability propagation regime, and meanwhile makes the
stable synfire propagation regime move to the lower right of
the $(K,p)$ panel.

\begin{figure}[!t]
\centering \includegraphics[width=7.8cm]{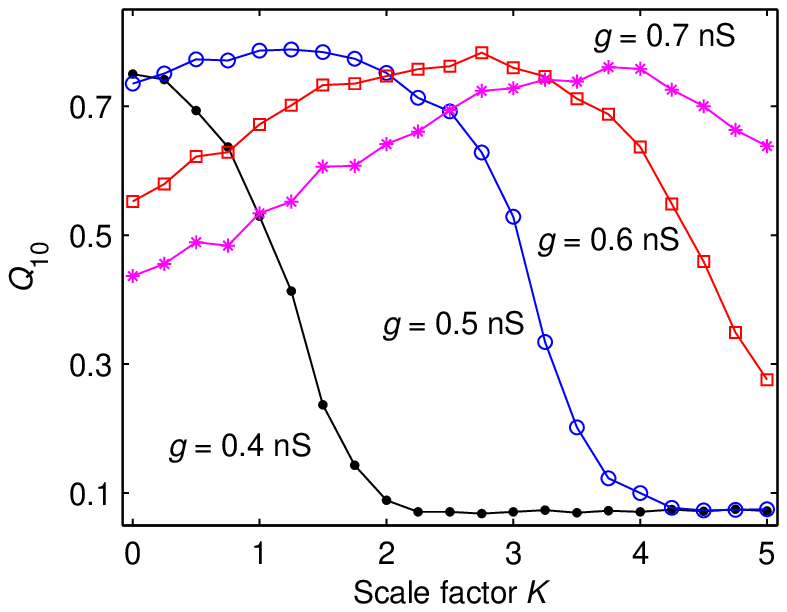}
\caption{\label{fig:16} (Color online) Effect of inhibition on
firing rate propagation. Here we show the value of $Q_{10}$ as
a function of scale factor $K$ for different excitatory
synaptic strengths. System parameters are $p=0.2$, and $D_s=D_t=0.6$
in all cases. Each data point is calculated based on 50 different
independent simulations with different random seeds.}
\end{figure}

For the case of firing rate propagation, we plot the value of
$Q_{10}$ versus the scale factor $K$ for different excitatory
synaptic strengths in Fig.~\ref{fig:16}, with a fixed
successful transmission probability $p=0.2$. When the
excitatory synaptic strength is small (for instance $g=0.4$
nS), due to weak excitatory synaptic interaction between
neurons the UREI feedforward neuronal network cannot transmit
the firing rate sufficiently even for $K=0$. In this case, less
and less neural information can be propagated to the final
layer of the considered network with the increasing of $K$.
Therefore, $Q_{10}$ monotonically decreases with the scale
factor $K$ at first and finally approaches to a low steady
state value. Note that here the low steady state value is
purely induced by the spontaneous neural firing activities,
which are caused by the additive Gaussian white noise. As the
excitatory synaptic strength grows, more neural information can
be successfully transmitted for small value of $K$. When $g$ is
increased to a rather large value, such as $g=0.6$ nS, the
coupling is so strong that too small scale factor will lead to
the excessive propagation of firing rate. However, in this
case, the propagation of firing rate can still be suppressed
provided that the relative strength of inhibitory and
excitatory synapses is strong enough. As a result, there always
exists an optimal scale factor to best support the firing rate
propagation for each large excitatory synaptic strength (see
Fig.~\ref{fig:16}). If we fix the value of $g$ (not too small),
then the similar results can also be observed by changing the
scale factor for a large successful transmission probability
(data not shown). Once again, this is due to the fact that
increasing $g$ and fixing $p$ is equivalent to increasing $p$
and fixing $g$ to a certain degree.

\section{Conclusion and discussion}
\label{sec:4}

The feedforward neuronal network provides us an effective way
to examine the neural activity propagation through multiple
brain regions. Although biological experiments suggested that
the communication between neurons is more or less unreliable
(Abeles 1991; Raastad et al. 1992; Smetters and Zador 1996), so
far most relevant computational studies only considered that
neurons transmit spikes based on reliable synaptic models. In
contrast to these previous work, we took a different approach
in this work. Here we first built a 10-layer feedforward
neuronal network by using purely excitatory neurons, which are
connected with unreliable synapses in an all-to-all coupling
fashion, that is, the so-called URE feedforward neuronal
network in this paper. The goal of this work was to explore the
dependence of both the synfire propagation and firing rate
propagation on unreliable synapses in the URE neuronal network,
but was not limited this type of feedforward neuronal network.
Our modelling methodology was motivated by experimental results
showing the probabilistic transmitter release of biological
synapses (Branco and Staras 2009; Katz 1966; Katz 1969;
Trommersh\"{a}user et al. 1999).

In the study of synfire mode, it was observed that the synfire
propagation can be divided into three types (i.e., the failed
synfire propagation, the stable synfire propagation, and the
synfire instability propagation) depending on whether the
synfire packet can be successfully and stably transmitted to
the final layer of the considered network. We found that the
stable synfire propagation only occurs in the suitable region
of system parameters (such as the successful transmission
probability and excitatory synaptic strength). For system
parameters within the stable synfire propagation regime, it was
found that both high synaptic reliability and strong excitatory
synaptic strength are able to support the synfire propagation
in feedforward neuronal networks with better performance and
faster transmission speed. Further simulation results indicated
that the performance of synfire packet in deep layers is mainly
influenced by the intrinsic parameters of the considered
network but not the parameters of the initial spike packet,
although the initial spike packet determines whether the
synfire propagation can be evoked to a great degree. In fact,
this is a signature that the synfire activity is governed by a
stable attractor, which is in agreement with the results given
in (Diesmann et al. 1999; Diesmann et al. 2001; Diesmann 2002;
Gewaltig et al. 2001).

In the study of firing rate propagation, our simulation results
demonstrated that both the successful transmission probability
and the excitatory synaptic strength are critical for firing
rate propagation. Too small successful transmission probability
or too weak excitatory synaptic strength results in the
insufficient firing rate propagation, whereas too large
successful transmission probability or too strong excitatory
synaptic strength has the tendency to lead to the excessive
propagation of firing rate. Theoretically speaking, better
tuning of these two synaptic parameters can help neurons encode
the neural information more accurately.

On the other hand, neuronal noise is ubiquitous in the brain.
There are many examples confirmed that noise is able to induce
many counterintuitive phenomena, such as stochastic resonance
(Collins et al. 1995a; Collins et al. 1995b; Collins et al.
1996; Chialvo et al. 1997; Guo and Li 2009) and coherence
resonance (Pikovsky and Kurths 1996; Lindner and
Schimansky-Geier 2002; Guo and Li 2009). Here we also
investigated how the noise influences the performance of signal
propagation in URE feedforward neuronal networks. The numerical
simulations revealed that noise tends to reduce the performance
of synfire propagation because it makes neurons desynchronized
and causes some spontaneous neural firing activities. Further
studies demonstrated that the survival rate of synfire
propagation is also largely influenced by the noise. In
contrast to the synfire propagation, our simulation results
showed that noise with appropriate intensity is able to enhance
the performance of firing rate propagation in URE feedforward
neuronal networks. In essence, it is because suitable noise can
help neurons in each layer maintain more accurate temporal
structural information of the the external input signal. Note
that the relevant mechanisms about noise have also been
discussed in several previous work (van Rossum et al. 2002;
Masuda and Aihara 2002; Masuda and Aihara 2003), and our
results are consistent with the findings given in these work.

Furthermore, we have also investigated the stochastic effect of
neurotransmitter release on the performance of signal
propagation in the URE feedforward neuronal networks. For both
the synfire propagation and firing rate propagation, we found
that the URE feedforward neuronal networks might display
different propagation performance, even when their average
amount of received neurotransmitter for each neuron in a time
instant remains unchanged. This is because the unreliability of
neurotransmitter release will add randomness to the system.
Different synaptic transmission probabilities will introduce
different levels of stochastic effect, and thus might lead to
different spiking dynamics and propagation performance. These
findings revealed that the signal propagation dynamics in
feedforward neuronal networks with unreliable synapses is also
largely influenced by the stochastic effect of neurotransmitter
release.

Finally, two supplemental work has been also performed in
this paper. In the first work, we compared both the synfire
propagation and firing rate propagation in URE feedforward
neuronal networks with the results in corresponding feedforward
neuronal networks composed of purely excitatory neurons but
connected with reliable synapses in an random coupling fashion
(RRE feedforward neuronal network). Our simulations showed that
several different results exist for both the synfire
propagation and firing rate propagation in these two different
feedforward neuronal network models. These results tell us that
we should be cautioned when using random connections to replace
unreliable synapses in modelling research. In the second work,
we extended our results to more generalized feedforward
neuronal networks, which consist not only of the excitatory
neurons but also of inhibitory neurons (UREI feedforward
neuronal network). The simulation results demonstrated that
inhibition also plays an important role in both types of neural
activity propagation, and better choosing of the relative
strength of inhibitory and excitatory synapses can enhance the
transmission capability of the considered network.

The results presented in this work might be more realistic than
those obtained based on reliable synaptic models. This is
because the communication between biological neurons indeed
displays the unreliable properties. In real neural systems,
neurons may make full use of the characteristics of unreliable
synapses to transmit neural information in an adaptive way,
that is, switching between different signal propagation modes
freely as required. Further work on this topic includes at
least the following two aspects: (i) since all our results are
derived from numerical simulations, an analytic description of
the synfire propagation and firing rate propagation in our
considered feedforward neuronal networks requires
investigation. (ii) Intensive studies on signal propagation in
the feedforward neuronal network with other types of
connectivity, such as the Mexican-hat-type connectivity
(Hamaguchi et al. 2004; Hamaguchi and Aihara 2005) and the
Gaussian-type connectivity (van Rossum et al. 2002), as well as
in the feedforward neuronal network imbedded into a recurrent
network (Aviel et al. 2003; Vogels and Abbott 2005; Kumar et
al. 2008), from the unreliable synapses point of view are
needed as well.

\section*{Acknowledgement}
We thank Feng Chen, Yuke Li, Qiuyuan Miao, Xin Wei and Qunxian
Zheng for valuable discussions on an early version of this
manuscript. We gratefully acknowledge the anonymous reviewers for
providing useful comments and suggestion, which greatly improved
our paper. We also sincerely thank one reviewer for reminding us
of a critical citation (Trommersh\"{a}user and Diesmann 2001).
This work is supposed by the National Natural Science Foundation
of China (Grant No. 60871094), the Foundation for the Author of
National Excellent Doctoral Dissertation of PR China, and the
Fundamental Research Funds for the Central Universities (Grant No.
1A5000-172210126). Daqing Guo would also like to thank the award
of the ongoing best PhD thesis support from the University of
Electronic Science and Technology of China.

\end{document}